ENVIRONMENTAL PROTECTION

# The new world atlas of artificial night sky brightness

Fabio Falchi,[1]* Pierantonio Cinzano,[1] Dan Duriscoe,[2] Christopher C. M. Kyba,[3,4] Christopher D. Elvidge,[5] Kimberly Baugh,[6] Boris A. Portnov,[7] Nataliya A. Rybnikova,[7] Riccardo Furgoni[1,8]

Artificial lights raise night sky luminance, creating the most visible effect of light pollution—artificial skyglow. Despite the increasing interest among scientists in fields such as ecology, astronomy, health care, and land-use planning, light pollution lacks a current quantification of its magnitude on a global scale. To overcome this, we present the world atlas of artificial sky luminance, computed with our light pollution propagation software using new high-resolution satellite data and new precision sky brightness measurements. This atlas shows that more than 80% of the world and more than 99% of the U.S. and European populations live under light-polluted skies. The Milky Way is hidden from more than one-third of humanity, including 60% of Europeans and nearly 80% of North Americans. Moreover, 23% of the world's land surfaces between 75°N and 60°S, 88% of Europe, and almost half of the United States experience light-polluted nights.

## INTRODUCTION

Light pollution is the alteration of night natural lighting levels caused by anthropogenic sources of light (1). Natural lighting levels are governed by natural celestial sources, mainly the Moon, natural atmospheric emission (airglow), the stars and the Milky Way, and zodiacal light. During moonless nights, the luminance of the clear sky background far from the Milky Way and zodiacal light is about 22 magnitude per square arcsecond (mag/arcsec$^2$) in the Johnson-Cousins V-band (2), equivalent to $1.7 \times 10^{-4}$ cd/m$^2$. Artificial light scattered in the atmosphere raises night sky luminance, creating the most visible negative effect of light pollution—artificial skyglow. In addition to hindering ground-based optical astronomical observations, the artificial brightening of the night sky represents a profound alteration of a fundamental human experience—the opportunity for each person to view and ponder the night sky. Even small increases in sky brightness degrade this experience. Light pollution is no longer only a matter for professional astronomers (3, 4). Although researchers from disparate fields are now interested in light pollution, its magnitude is poorly known on a global scale because measurements are sporadically distributed across the globe. To overcome this, we present the world atlas of artificial sky luminance, which was obtained with our dedicated light pollution propagation software using the new calibrated, high–dynamic range, high-resolution data from the Visible Infrared Imaging Radiometer Suite (VIIRS) Day/Night Band (DNB), new precision charge-coupled device (CCD) brightness measurements, and a new database of Sky Quality Meter (SQM) measurements.

Light pollution is one of the most pervasive forms of environmental alteration (5). It affects even otherwise pristine sites because it is easily observed during the night hundreds of kilometres from its source in landscapes that seem untouched by humans during the day (6), damaging the nighttime landscapes even in protected areas, such as national parks (for example, the light domes of Las Vegas and Los Angeles as seen from Death Valley National Park). Notwithstanding its global presence, light pollution has received relatively little attention from environmental scientists in the past. This is changing, as attested by the rapidly increasing rate of published works on the subject.

The atlas we present here is intended to help researchers in all fields who may be interested in the levels of light pollution for their studies (for example, in astronomy, ecology, environmental protection, and economics).

## RESULTS

### Upward function and maps

Using the maximum likelihood fit described in Materials and Methods, we found an average upward emission function that best fits the whole data set (red curve in Fig. 1). This upward function is not meant to be considered a "real" or "best" upward function but is simply the function that produces the best statistical fit to the entire observational data set. Factors other than the actual light intensity distribution may influence its shape (for example, atmospheric transparency that is higher or lower than that assumed by the model). The fit suggests that, in addition to the Lambertian distribution resulting from surface reflections, low-angle upward emissions are an important component of light emission from cities. This component presumably originates from poorly shielded luminaires. The fact that the main component of the upward flux was found to be the Lambertian one does not mean that the reflected light is the origin of the main component of the artificial sky brightness. In fact, as previously demonstrated (7, 8), the sky brightness outside cities is dominated by the component of the light escaping at low angles above the horizon plane, exactly where the fit upward function differs most from the pure Lambertian distribution.

Maps were produced to show the zenith artificial sky brightness in twofold increasing steps as a ratio to the natural sky brightness (Figs.

[1]Istituto di Scienza e Tecnologia dell'Inquinamento Luminoso (Light Pollution Science and Technology Institute), 36016 Thiene, Italy. [2]National Park Service, U.S. Department of Interior, Natural Sounds and Night Skies Division, Fort Collins, CO 80525, USA. [3]Deutsches GeoForschungsZentrum GFZ, Potsdam, Germany. [4]Leibniz–Institute of Freshwater Ecology and Inland Fisheries, Berlin, Germany. [5]Earth Observation Group, National Oceanic and Atmospheric Administration's National Centers for Environmental Information, Boulder, CO 80305, USA. [6]Cooperative Institute for Research in Environmental Sciences, University of Colorado, Boulder, CO 80309, USA. [7]Department of Natural Resources and Environmental Management, Faculty of Management, University of Haifa, 3498838 Haifa, Israel. [8]American Association of Variable Star Observers, Cambridge, MA, USA.
*Corresponding author. Email: falchi@istil.it

2 to 8). The maps were calibrated to match the time of satellite overpass, at around 1 a.m. Because of the decrease in artificial illumination during the night, brighter skies should typically be expected for observations made earlier in the night. We chose 22.0 mag/arcsec$^2$, corresponding to 174 μcd/m$^2$, as a typical brightness of the night sky background during solar minimum activity, excluding stars brighter than magnitude 7, away from Milky Way and from Gegenschein and zodiacal light. Natural airglow variations, even during the same night, can cause more than half a magnitude variation in the background sky brightness at unpolluted sites. Measurements of the sky brightness made with wide-field instruments that integrate the light arriving from a substantial portion of the sky (for example, SQM and SQM-L) include the light from naked eye stars, increasing the detected sky brightness. If this were not taken into account, it could bias a comparison with atlas predictions. Table 1 shows the color levels used for the maps.

For the purpose of this atlas, we set the level of artificial brightness under which a sky can be considered "pristine" at 1% of the natural background. Although 1% (1.7 μcd/m$^2$) is a nearly unmeasurable incremental effect at the zenith (usually the darkest part of the sky hemisphere), it is generally much larger near the horizon in the direction of the source(s). For areas protected for scenic or wilderness character, this horizon glow has a significant impact on the values of solitude and the absence of visual intrusion of human development.

The dark gray level (1 to 2%) sets the point where attention should be given to protect a site from a future increase in light pollution. Blue (8 to 16%) indicates the approximate level where the sky can be considered polluted on an astronomical point of view, as indicated by recommendation 1 of IAU Commission 50 (9). The winter Milky Way (fainter than its summer counterpart) cannot be observed from sites coded in yellow, whereas the orange level sets the point of artificial brightness that masks the summer Milky Way as well. This level corresponds to an approximate total sky brightness of between 20.6 and 20.0 mag/arcsec$^2$ (0.6 to 1.1 mcd/m$^2$). With this sky brightness, the summer Milky Way in Cygnus may be only faintly detectable as a small increase in the sky background luminosity. The Sagittarius Star Cloud is the only section of the Milky Way that is still visible at this level of pollution when it is overhead, as observed from southern latitudes. Red indicates the approximate threshold where Commission Internationale de l'Eclairage (10) puts the transition between scotopic vision and mesopic vision (1 mcd/m$^2$). Also inside the range of the red level, the sky has the same luminosity as a pristine uncontaminated sky at the end of nautical twilight (1.4 mcd/m$^2$) (*11*). This means that, in places with this level of pollution, people never experience conditions resembling a true night because it is masked by an artificial twilight.

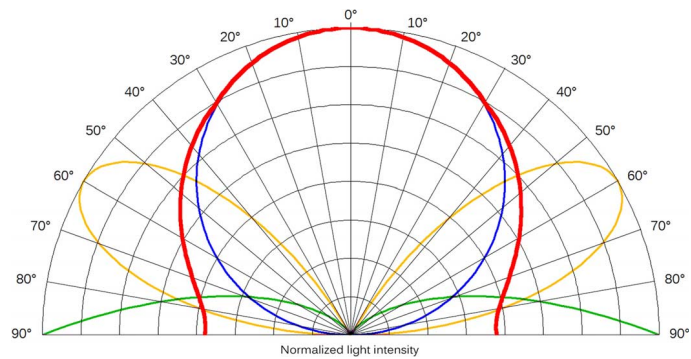

**Fig. 1. Upward emission functions used to compute the maps.** The polar graph shows the three different light intensity distributions used to compute the three map versions: the Lambertian distribution with a peak toward the zenith (map *A*; blue), the function with peak intensity at low angles above the horizon plane (map *B*; green), and the function with peak at intermediate angles, 30° above the horizon plane (map *C*; yellow). The thick red line shows the overall best-fitting function.

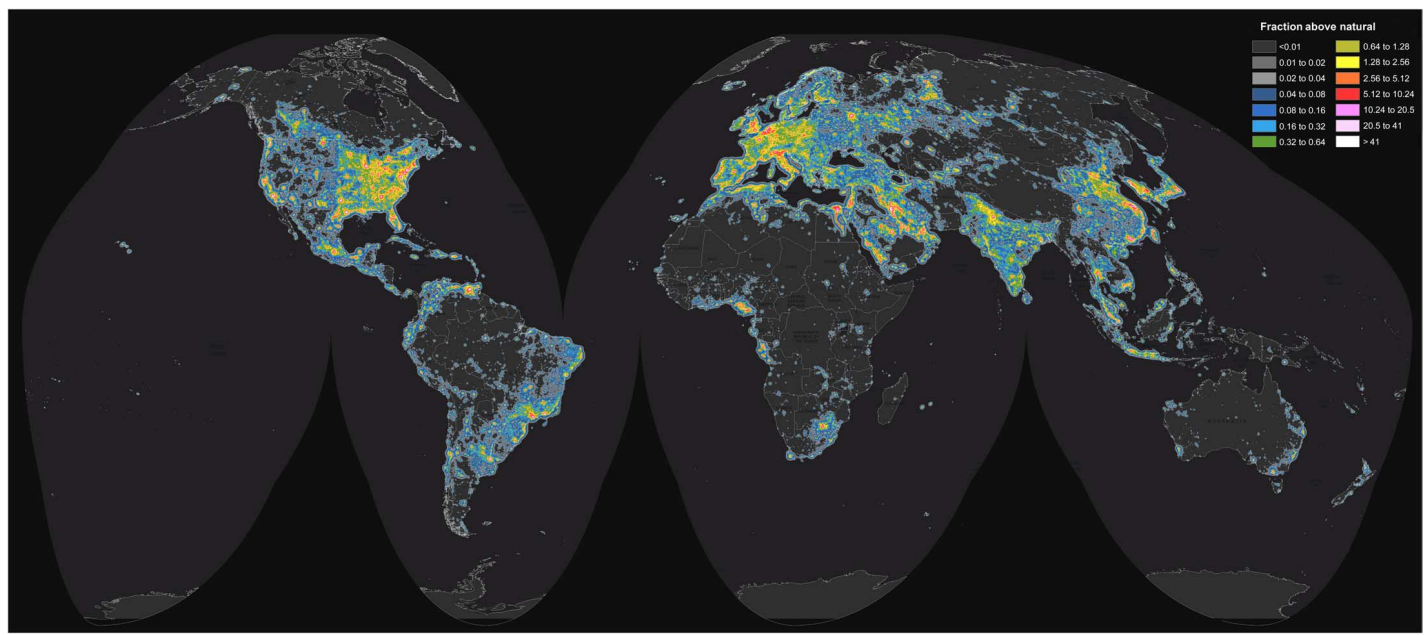

**Fig. 2. World map of artificial sky brightness.** The map shows, in twofold increasing steps, the artificial sky brightness as a ratio to the natural sky brightness (assumed to be 174 μcd/m$^2$). Table 1 indicates the meaning of each color level.

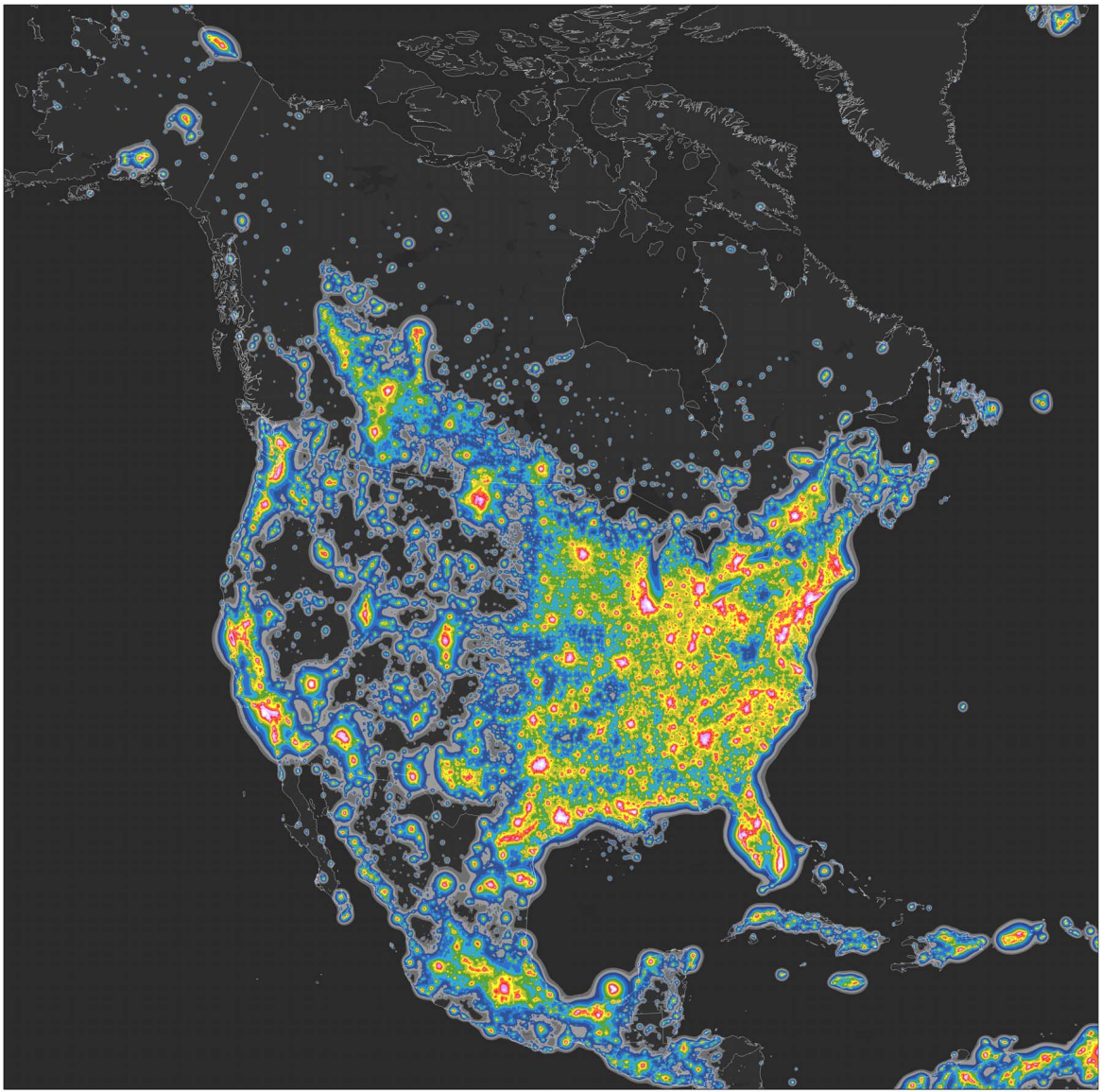

**Fig. 3. Map of North America's artificial sky brightness, in twofold increasing steps, as a ratio to the natural sky brightness.** Table 1 indicates the meaning of each color level.

A geographic proximity analysis reveals locations on Earth where residents would have to travel very long distances to reach a land-based observing site of sufficient sky quality where certain features of the night sky are revealed. The location on Earth that is most distant from having the possibility to get a hint of a view of the Milky Way (artificial sky brightness at zenith <688 μcd/m$^2$) is an area near Cairo, Egypt, in the Nile Delta region. The other widest regions where the Milky Way is no longer visible include the Belgium/Netherlands/Germany (Dortmund to Bonn cities) transnational region, the Padana plain in northern Italy, and the Boston to Washington series of cities in the northeastern United States. Other large areas where the Milky Way is lost are the London to Liverpool/Leeds region in England, and regions surrounding Beijing and Hong Kong in China and Taiwan. People living near Paris would have to travel 900 km to Corsica, Central Scotland, or Cuenca province, Spain, to find large territories where the zenith is essentially unaffected by light pollution (artificial sky brightness <8% of the natural background); even in these places, significant skyglow would be present near the horizon. The pristine sky (artificial sky brightness <1% of the natural background) nearest to Neuchâtel, Switzerland, is more than 1360 km away, in northwestern Scotland, Algeria, or Ukraine (see Fig. 9). There are islands that are farther yet to land-based pristine skies. More than 1400 km of travel is needed from Bermuda to the nearest pristine sky in Nova Scotia. The land on Earth that is farthest from a pristine sky is in Azores, more than 1750 km away from the pristine sky of the western Sahara.

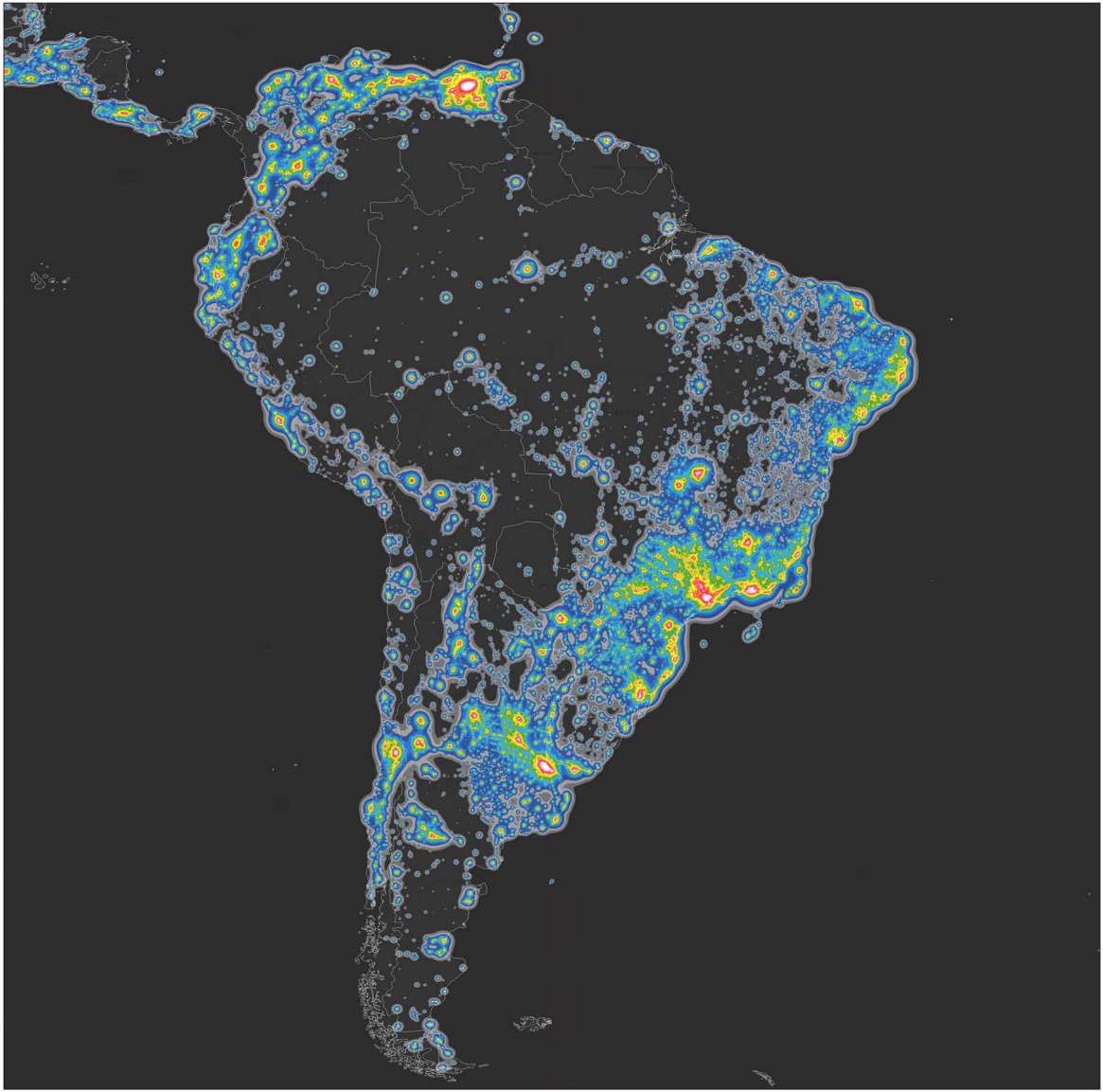

**Fig. 4. Map of South America's artificial sky brightness, in twofold increasing steps, as a ratio to the natural sky brightness.** Table 1 indicates the meaning of each color level.

### Findings: Statistics of nighttime brightness

This atlas shows, for the first time in 15 years, the impact of the pollution generated by artificial night lights on world population. Figure 10 presents the data of the world atlas with the levels used to compute the statistics to show the visual impacts of the different levels of light pollution.

We found that about 83% of the world's population and more than 99% of the U.S. and European populations live under light-polluted skies (that is, where artificial sky brightness at the zenith is >14 $\mu cd/m^2$). Due to light pollution, the Milky Way is not visible to more than one-third of humanity, including 60% of Europeans and nearly 80% of North Americans. Moreover, 23% of the world's land surfaces between 75°N and 60°S, 88% of Europe, and almost half of the United States experience light-polluted nights.

The countries with the populations least affected by light pollution are Chad, Central African Republic, and Madagascar, with more than three-quarters of their inhabitants living under pristine sky conditions. The countries and territories with the largest nonpolluted areas are Greenland (only 0.12% of its area does not have pristine skies), Central African Republic (0.29), Niue (0.45%), Somalia (1.2%), and Mauritania (1.4%).

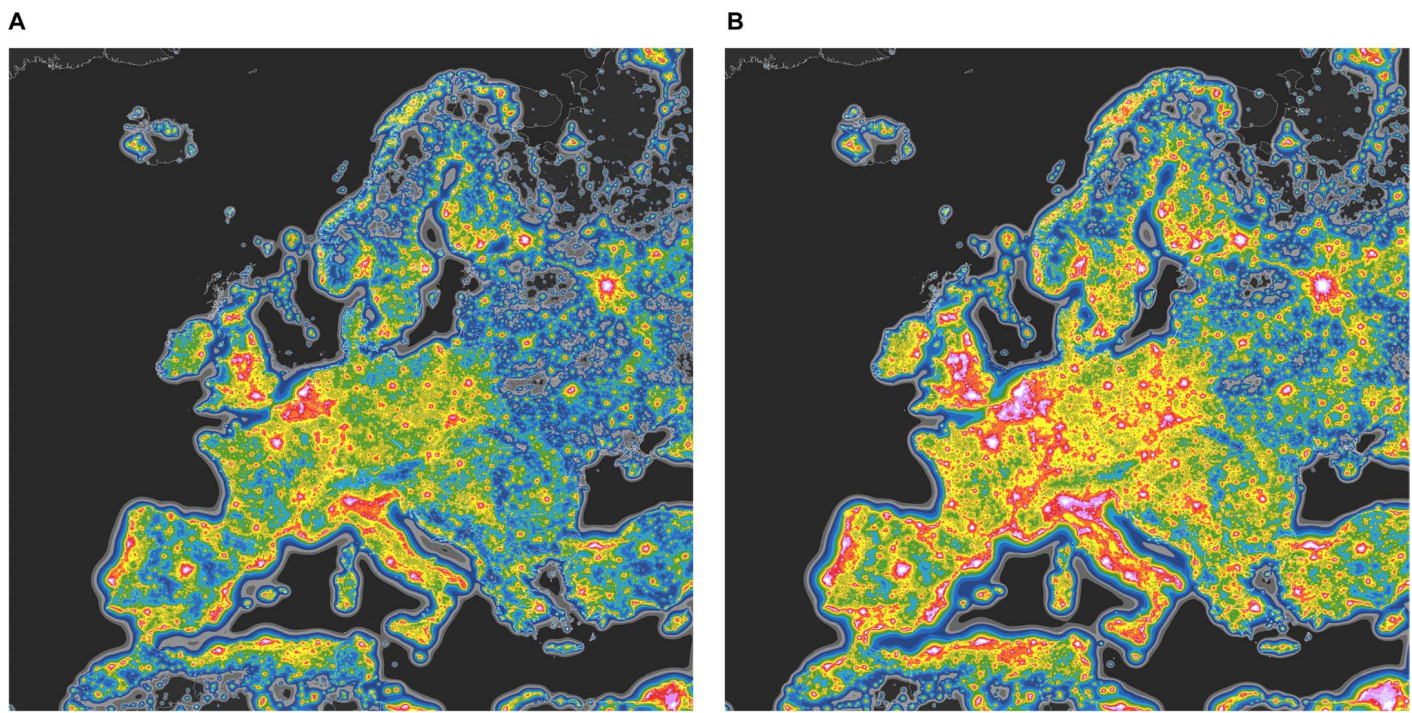

**Fig. 5. Maps of Europe's artificial sky brightness, in twofold increasing steps, as a ratio to the natural sky brightness (assumed to be 174 μcd/m².** (**A**) The map shows the artificial sky brightness in V-band, as in the other maps. (**B**) The map shows the forecast of the perceived sky brightness for a dark-adapted eye after a transition toward 4000K CCT LED technology, without increasing the photopic flux of currently installed lamps. Table 1 indicates the meaning of each color level.

On the other side, the most light-polluted country is Singapore, where the entire population lives under skies so bright that the eye cannot fully dark-adapt to night vision. Other populations experiencing this level of light pollution are Kuwait (98%), Qatar (97%), United Arab Emirates (93%), Saudi Arabia (83%), South Korea (66%), Israel (61%), Argentina (58%), Libya (53%), and Trinidad and Tobago (50%); all of these countries have more than half of their inhabitants living under extremely bright skies.

The possibility of seeing the Milky Way from home is precluded to all of Singapore, San Marino, Kuwait, Qatar, and Malta inhabitants, and for 99%, 98%, and 97% of the population of United Arab Emirates, Israel, and Egypt, respectively. The countries with the largest part of their territory where the Milky Way is hidden by light pollution are Singapore and San Marino (100%), Malta (89%), West Bank (61%), Qatar (55%), Belgium and Kuwait (51%), Trinidad and Tobago and the Netherlands (43%), and Israel (42%).

Among the G20 countries, Saudi Arabia and South Korea have the highest percentage of the population exposed to extremely bright skies, whereas Germany is the least polluted using this same measure. The territories of Italy and South Korea are the most polluted among the G20 countries, whereas Australia is the least polluted. As a reminder, the results presented here are computed for 1 a.m. observations to match satellite overpasses, implying that the situation in the early evening is even worse.

Figure 11 shows the population statistics for the G20 countries. Figure 12 shows the area statistics for the G20 countries. Figures 13 and 14 show the population statistics for the 20 least polluted countries and the 20 most polluted countries. The complete results of the analysis are reported in Table 2.

## DISCUSSION

The atlas has been computed using several constant assumptions, including the transparency of the atmosphere, the upward emission function of cities, the spectrum of artificial lights, and the hour of the night of the observation. The more the actual conditions differ from these assumptions, the greater the deviation in artificial light will be compared to the atlas prediction. In particular, observation sites near oil-well flares (for example, in North Dakota, North Sea, Sahara, and elsewhere) or volcanoes (for example, Kilauea volcano on Hawaii Big Island) are predicted to be far brighter than our eyes would perceive the sky mainly because of the near-infrared sensitivity of the VIIRS and secondarily because of the different upward emission function. Other sources of differences between prediction and actual measurements may be as follows: snow coverage, different outdoor lighting habits (for example, effective laws against light pollution), the presence of atypical lights (for example, greenhouse lighting, fishing lights, and gas flares), curfew, the presence of temporary lights detected by satellites that are no longer active, and vice versa.

Snow coverage acts by enhancing the upward flux coming from cities, and an almost linear relationship between ground reflectance and artificial sky brightness has been found in models inside and near (30 km) cities (*12*). Owing to the fact that snow is usually removed promptly from lit roads, the brightening effect of snow may be lower than expected. As a gauge for the size of this effect, 1.3- to 2.6-fold increases in sky brightness were measured when snow was partially present on roads (*13*). Snow coverage also reflects the artificial light coming from the polluted sky, increasing the artificial sky brightness by less than 10% (*13*).

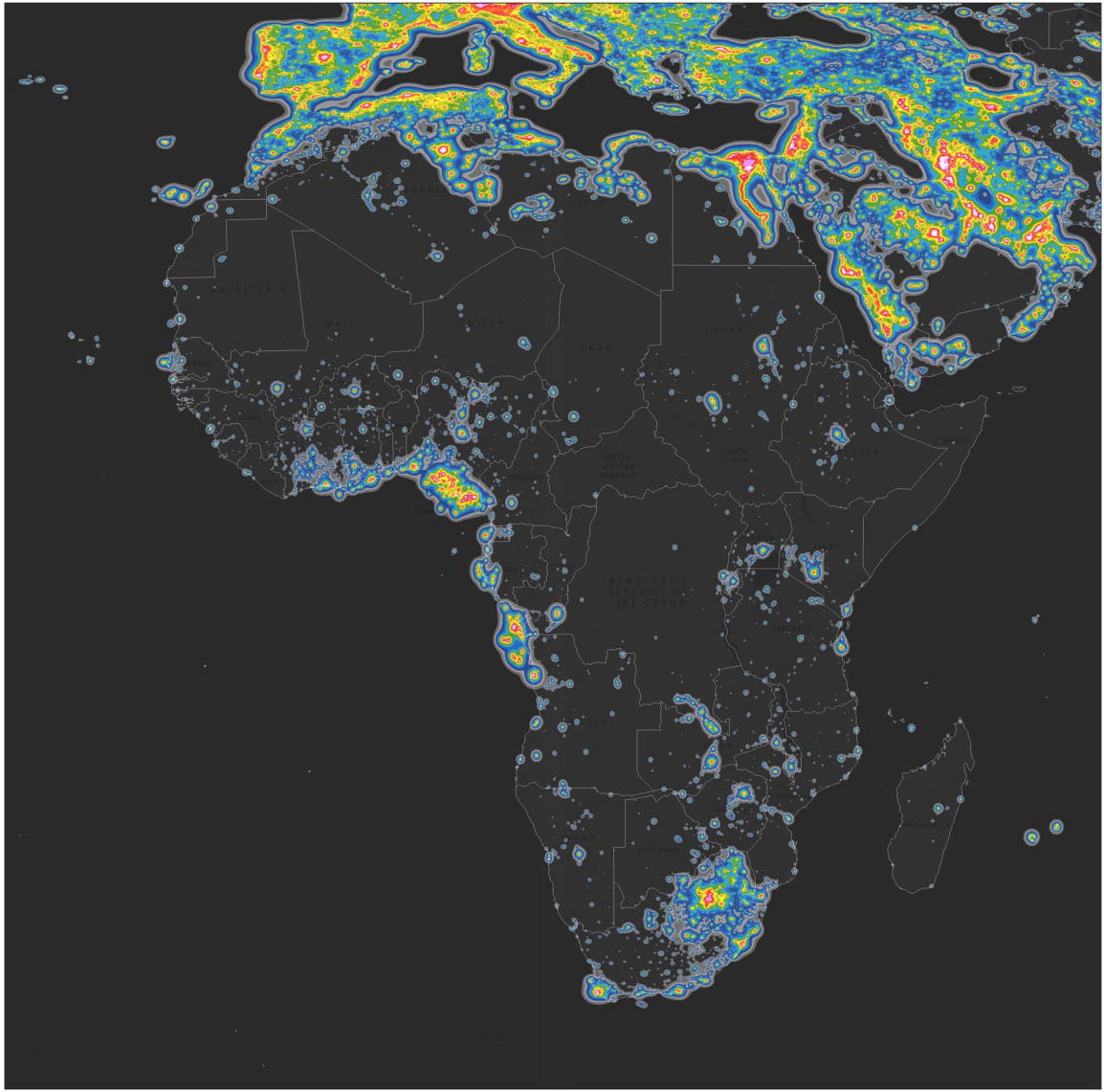

**Fig. 6. Map of Africa's artificial sky brightness, in twofold increasing steps, as a ratio to the natural sky brightness.** Table 1 indicates the meaning of each color level.

The atmospheric conditions are of paramount importance. If the sky is clearer or hazier than assumed, changes in sky brightness may be expected. A small increase in aerosols usually results in greater pollution near and inside cities, with the potential for darkening at sites far from light sources (very high levels of aerosols can have other effects). The relative positions of light sources and observation sites may result in a nonlinear behavior [for example, in Cinzano *et al.* (*14*)]. When the sky is overcast, a several-fold increase in skyglow is to be expected near cities [for example, 10 times more in Berlin (*15*)]. The ecological consequences of artificial night lighting are therefore larger than one might suppose simply by looking at the levels experienced during clear nights.

Comparisons with the first atlas should take into account the differences in the hour of the night of the satellite overpass and calibration data, the different upward functions used, the sea level–versus–altitude computations, the different snow coverage in some northern regions, the different assumed natural luminance levels, and the different levels of color code.

In Fig. 5B, we show an illustrative forecast map of Europe for the possible effect on scotopic perception of the sky for a transition from high-pressure sodium (HPS) lighting to white light-emitting diode (LED) lighting. Assuming that the photopic flux and upward emission function remain equal, a 4000K white LED light is about 2.5 times more polluting for the scotopic band of the spectrum than is HPS lighting (*16*–*18*). Higher correlated color temperature (CCT) sources are more polluting, whereas lower CCT sources are less. This implies that unless blue-light emission is restricted, a transition toward this technology can be expected to more than double the night sky brightness as perceived by our dark-adapted eyes. The map in Fig. 5B shows a 2.5-fold increase over the map in Fig. 5A. Other problems also arise when using high-CCT sources and white LEDs, such as longer time to recover dark adaption (*19*) and effects on many physiologic functions (*20*). This increase in the scotopic band and in the blue part of the spectrum will not be detected by VIIRS DNB because of its lack of sensitivity at wavelengths shorter than 500 nm. Because of this, the blue-light emission peak of white LEDs is not detected, and equal radiance values measured by the satellite do not correspond to equal brightness impressions for HPS and LED sources. This means that the "blue blindness" of the VIIRS DNB will falsely suggest a reduction in light pollution in many cities in the near future, whereas the brightness of the sky as seen by human eyes will in fact increase.

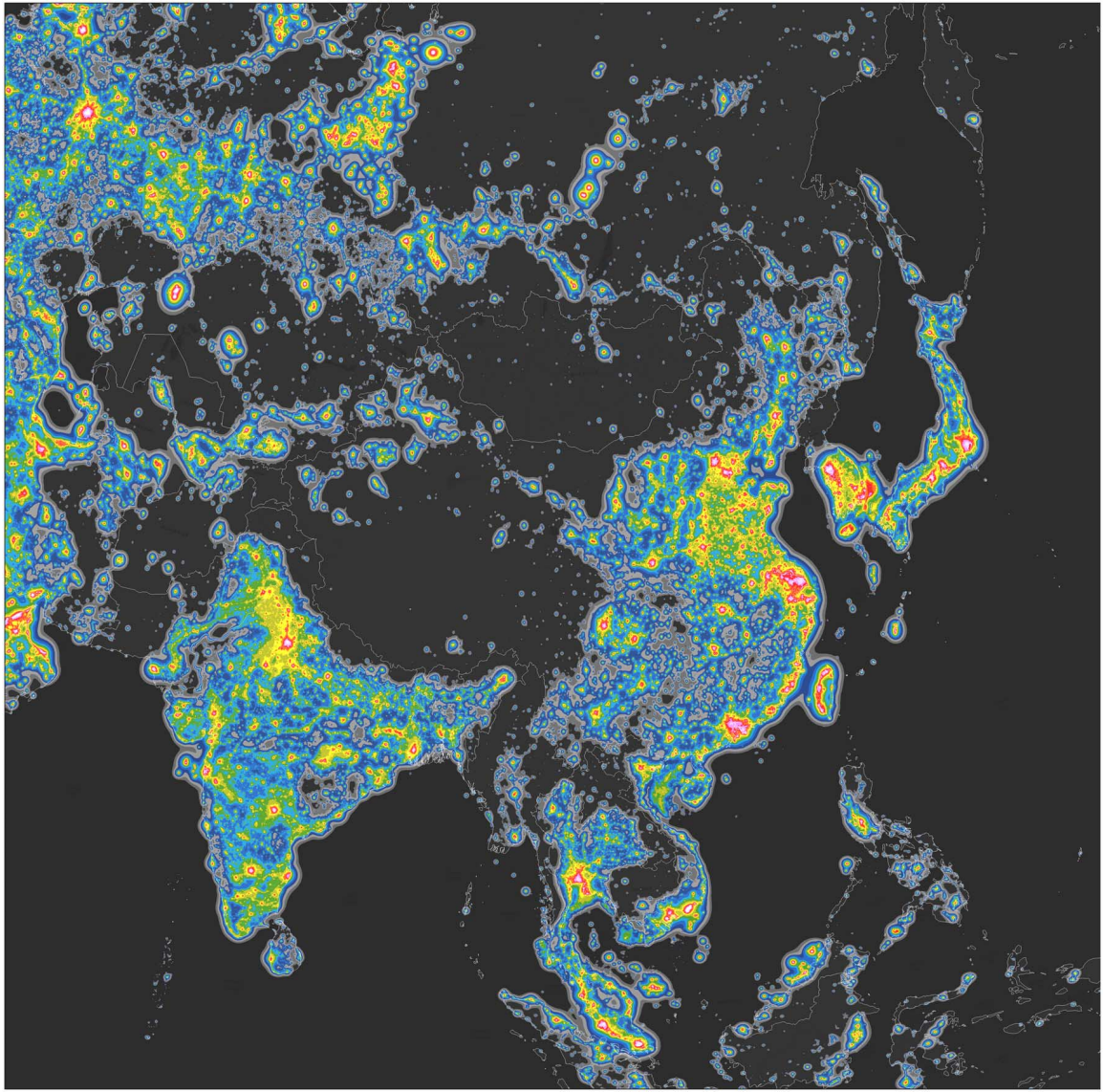

**Fig. 7. Map of Asia's artificial sky brightness, in twofold increasing steps, as a ratio to the natural sky brightness.** Table 1 indicates the meaning of each color level.

## CONCLUSIONS

The results presented here demonstrate that light pollution is a global issue. Most of the world is affected by this problem, and humanity has enveloped our planet in a luminous fog that prevents most of Earth's population from having the opportunity to observe our galaxy. This has a consequent potential impact on culture that is of unprecedented magnitude. Moreover, light pollution causes global ecological consequences (*21*), poses public health issues (*22*–*24*), and wastes energy and money (*25*). Light pollution needs to be addressed immediately because, even though it can be instantly mitigated (by turning off lights), its consequences cannot (for example, loss of biodiversity and culture). Fortunately, techniques to substantially reduce light pollution are already known (*16*), and some of them have already been implemented at a relatively large scale (for example, Lombardia and most other Italian regions, Slovenia, two regions in Chile, and part of Canary Islands).

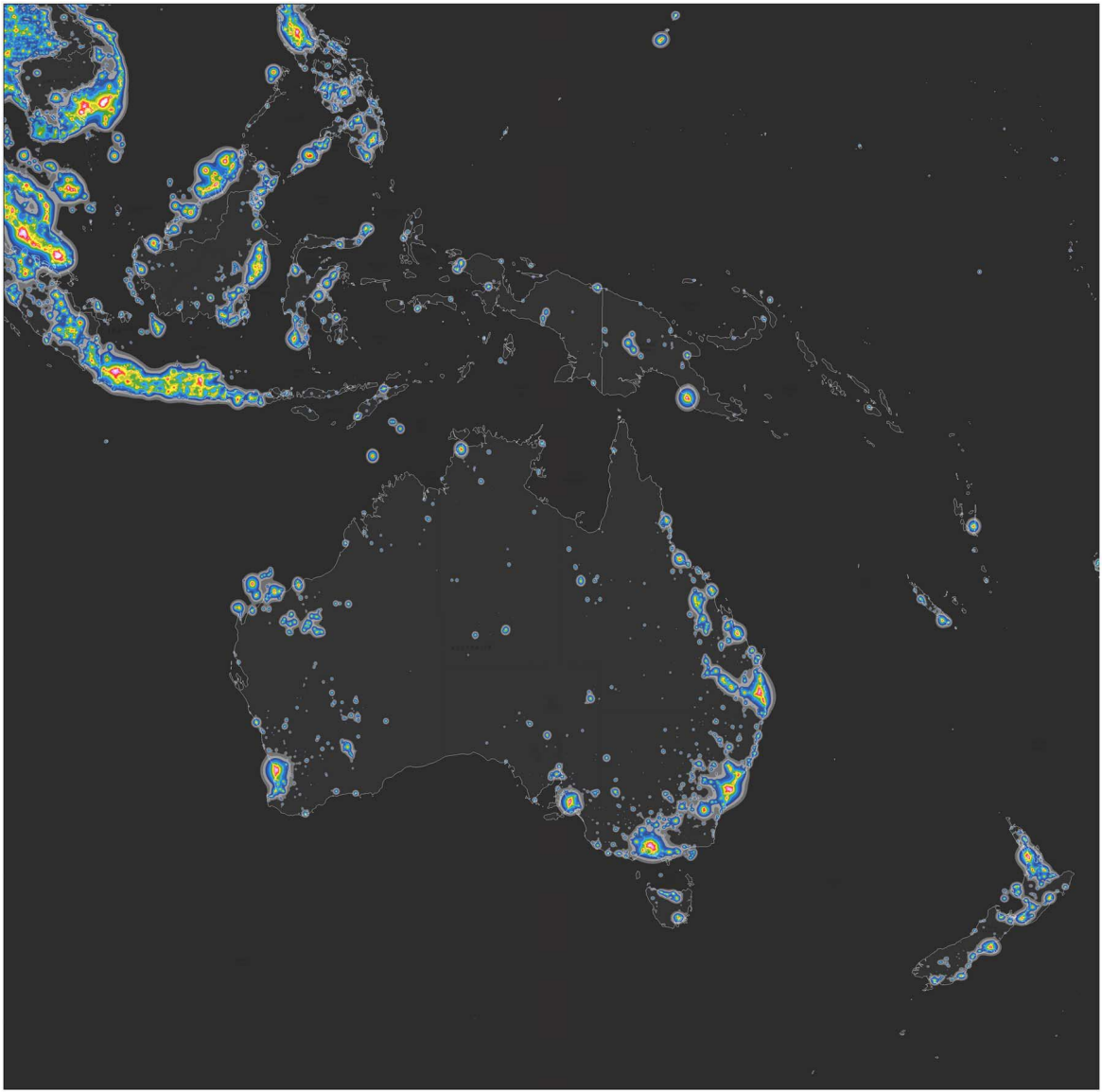

**Fig. 8. Map of artificial sky brightness for Australia, Indonesia, and New Zealand, in twofold increasing steps, as a ratio to the natural sky brightness.** Table 1 indicates the meaning of each color level.

The main prescriptions to lower light pollution are as follows: full shielding of lights (that is, do not allow luminaires to directly send any light at and above the horizon or outside the area to be lit), using the minimum light for the task, shutting off light or lowering its levels substantially when the area is not in use, decreasing the total installed flux (as is happening to most other pollutants), and strongly limiting the "blue" light that interferes with circadian rhythms and scotopic vision.

Technology may help to further reduce the impact of light pollution with the implementation of adaptive lighting (for example, street lighting that is governed by real-time sensors of traffic and meteorological conditions, substantially reducing the light during most of the night, in periods of low or no traffic). Looking further toward the future, public street lighting would not be necessary for driverless cars.

**Table 1. Color levels used in the maps.** The first column gives the ratio between the artificial brightness and the natural background sky brightness (assumed to be 174 μcd/m$^2$); the second column gives the artificial brightness (μcd/m$^2$); the third column gives the approximate (that is, assuming a natural background of 22 mag/arcsec$^2$) total brightness (mcd/m$^2$); and the fourth and fifth columns give the colors.

| Ratio to natural brightness | Artificial brightness (μcd/m$^2$) | Approximate total brightness (mcd/m$^2$) | Color | |
|---|---|---|---|---|
| <0.01 | <1.74 | <0.176 | Black | |
| 0.01–0.02 | 1.74–3.48 | 0.176–0.177 | Dark gray | |
| 0.02–0.04 | 3.48–6.96 | 0.177–0.181 | Gray | |
| 0.04–0.08 | 6.96–13.9 | 0.181–0.188 | Dark blue | |
| 0.08–0.16 | 13.9–27.8 | 0.188–0.202 | Blue | |
| 0.16–0.32 | 27.8–55.7 | 0.202–0.230 | Light blue | |
| 0.32–0.64 | 55.7–111 | 0.230–0.285 | Dark green | |
| 0.64–1.28 | 111–223 | 0.285–0.397 | Green | |
| 1.28–2.56 | 223–445 | 0.397–0.619 | Yellow | |
| 2.56–5.12 | 445–890 | 0.619–1.065 | Orange | |
| 5.12–10.2 | 890–1780 | 1.07–1.96 | Red | |
| 10.2–20.5 | 1780–3560 | 1.96–3.74 | Magenta | |
| 20.5–41 | 3560–7130 | 3.74–7.30 | Pink | |
| >41 | >7130 | >7.30 | White | |

Light pollution is also a consequence of the belief that artificial light increases safety on roads and prevents crimes, but this belief is not based on scientific evidence (26, 27). In a time of limited resources, countries should carefully invest money in effective ways to solve problems. For this reason, randomized controlled trials should be used to study the effects—positive, negative, or null—of implementing lighting as a means to reduce crimes and road accidents. For example, it could be the case that drivers respond to increased visibility by driving faster, increasing the risk of accidents. In addition, street lighting is commonly mounted on poles, and poles are dangerous 24 hours a day. The net influence of street lighting, de facto, is still unknown.

It is possible to imagine two scenarios for the future. Perhaps the current generation will be the final generation to experience such a light-polluted world, as light pollution is successfully controlled. Alternatively, perhaps the world will continue to brighten, with nearly the entire population never experiencing a view of the stars, as in Isaac Asimov's *Nightfall* novel and short story.

## MATERIALS AND METHODS

Sky brightness modeling using measured upward radiance from artificial sources as input is a proven method for describing light pollution at individual sites (28–30), over a region (31, 32), or even over the entire Earth (5). Remote sensing of upward radiance, along with sky brightness modeling, is used as a substitute for sky luminance observations, which are available only at selected locations.

### Satellite data

The atlas takes advantage of the newly available, low-light imaging data from the VIIRS DNB sensor on the Suomi National Polar-orbiting Partnership (NPP) satellite. The DNB achieves nightly global coverage with a swath width of approximately 3000 km, with each pixel having a spatial resolution of 742 m. These data have almost 7 times better native linear resolution and 256 times better dynamic range compared to the Defense Meteorological Satellite Program Operational Linescan System (DMSP-OLS) data (33). The DNB has on-board calibration and is reported in radiance units (W cm$^{-2}$ sr$^{-1}$). The DNB is sensitive to light in the range 0.5 to 0.9 μm, so its sensitivity spans out into the near-infrared region, beyond the range of the human eye, whereas it leaves out the blue and violet parts of the visible spectrum. The preflight relative spectral response curve is shown in Fig. 15. Although an instrument with sensitivity more closely matched to the human eye is preferred, all currently available global low-light imaging data (DNB-OLS) are sensitive to near-infrared light and blind in the blue part of the visible spectrum. This will prevent a good control of the evolution of light pollution in this important spectral band, where the white LEDs now being installed have strong emissions.

To generate the atlas, a complete view of Earth at night is required. Because nighttime lights are dynamic in nature, can be obscured by clouds, and contain types of lighting not desired for use in the atlas (for example, fires, lightning, and aurora), a composite product is needed. The Earth Observation Group (EOG) at the National Oceanic and Atmospheric Administration's National Centers for Environmental Information has been developing methods to create global maps of nighttime lights from the swath-level DNB data. These composite products are generated as monthly 15–arcsec grids, with each grid cell representing an average DNB radiance value. Only cloud-free nighttime DNB data on nights without moonlight present are included in this average. In addition, DNB data affected by stray light and other sensor-specific artifacts are filtered out before averaging (*34*).

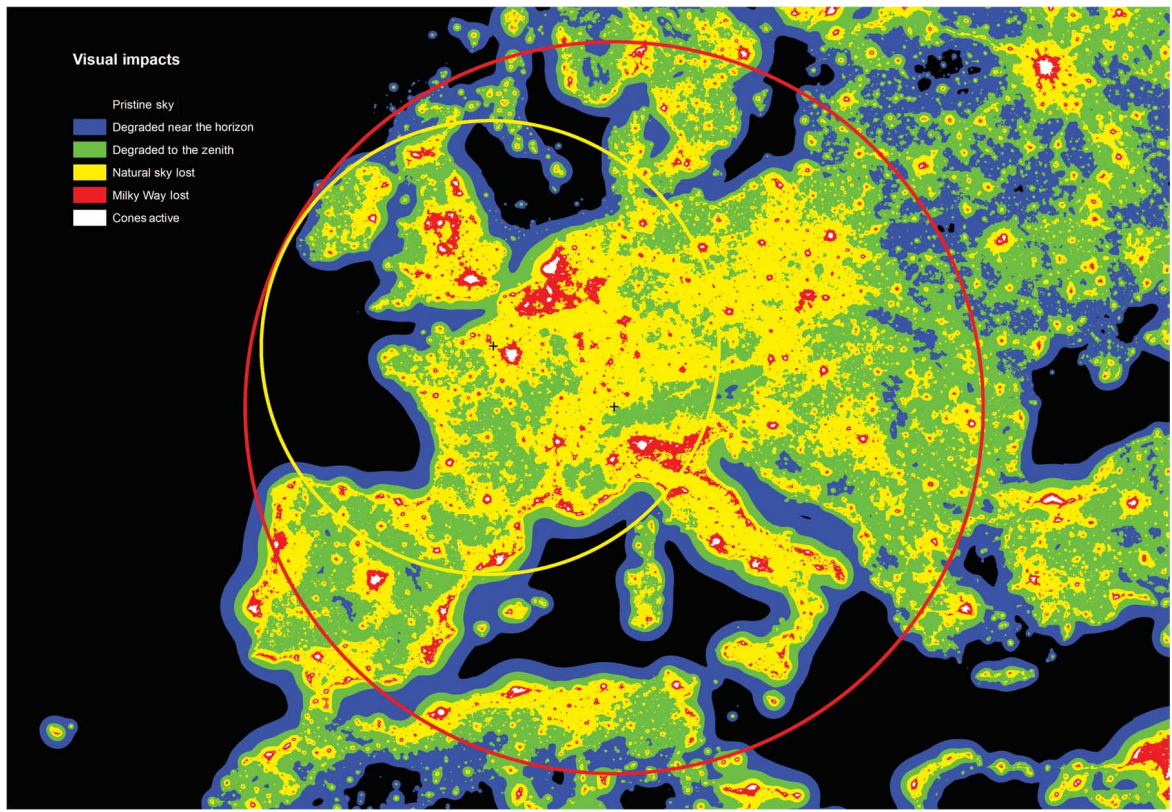

**Fig. 9. Places on Earth farthest from pristine skies and unpolluted zenith skies.** The sky brightness levels used here indicate the following: up to 1% above the natural light (0 to 1.7 μcd/m², black); from 1 to 8% above the natural light (1.7 to 14 μcd/m²; blue); from 8 to 50% above natural nighttime brightness (14 to 87 μcd/m²; green); from 50% above natural to the level of light under which the Milky Way is no longer visible (87 to 688 μcd/m²; yellow); from Milky Way loss to estimated cone stimulation (688 to 3000 μcd/m²; red); and very high nighttime light intensities, with no dark adaption for human eyes (>3000 μcd/m²; white). The circles indicate the distance toward arriving at large territories (that is, very small islands, such as Ile d'Ouessant, France, are not considered here), toward a sky relatively unpolluted at the zenith (yellow circle), and toward a pristine sky (red circle).

At the time of the creation of the atlas, the EOG had completed processing of 6 months of DNB data from 2014: May, June, September, October, November, and December. These data were combined into one composite product, with each grid cell representing an average radiance value for all observations in the 6-month period. Because the EOG was still developing algorithms to remove fires and other ephemeral lights from the DNB composites, the 2013 stable lights product from the DMSP-OLS sensor was used to mask out nonpersistent light sources. The use of November and December 2014 data means that radiance is somewhat increased in towns that had considerable snow cover at that time (for example, Edmonton and Calgary in Canada, but not Vancouver or Toronto, nor most northern cities in the United States).

**Mapping technique**
The Falchi *et al.* (*35*) maps were computed, following the method of Cinzano *et al.* (*36*), in Johnson-Cousins V-band, with a standard clear US62 atmosphere, an aerosol clarity (*37*) of $K = 1$ corresponding to a vertical extinction at sea level of $\Delta m = 0.33$ mag in the V-band, a horizontal visibility of $\Delta x = 26$ km, and an optical depth of $\tau = 0.31$. The vertical extinction at sea level becomes $\Delta m = 0.21$ mag for sites at 1000-m altitude, $\Delta m = 0.15$ mag for sites at 2000-m altitude, and $\Delta m = 0.11$ mag for sites at 3000-m altitude.

The zenith sky brightness at each site was obtained by integrating the contribution of the light arriving from sources up to a distance of 195 km, corresponding to a radius of 210 pixels at the equator. Maps were computed by taking into account the elevation above sea level of the sites, as given by GTOPO30 digital elevation data (*38*). Earth curvature screening—but not the screening effects of mountains—was considered. These effects are generally small, except for particular cases (*36*). Computing an atlas including the effects of mountain screening was not possible with the power of the computers at our disposal (36 Intel i5 PCs). The computation of the atlas required an equivalent of 200 days of time on an Intel i5 PC.

Although zenith brightness gives direct information for only one point in the sky (usually the darkest), it can be used to infer other physical quantities, such as horizontal illuminance (*39*). We are working on modeling hemispheric predictions for large territories to calculate other quantities (for example, horizontal illuminance, average and maximum vertical illuminance and luminance, and scalar illuminance).

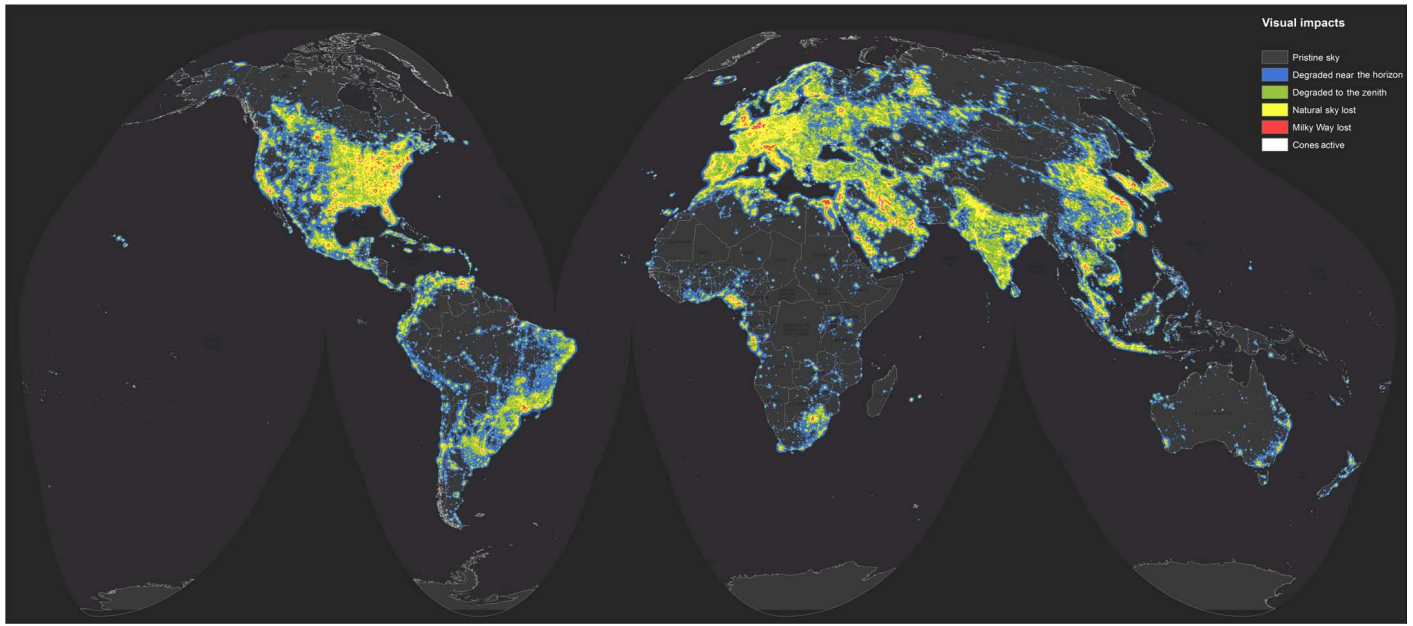

**Fig. 10. Map of light pollution's visual impact on the night sky.** The sky brightness levels are those used in the tables and indicate the following: up to 1% above the natural light (0 to 1.7 μcd/m$^2$; black); from 1 to 8% above the natural light (1.7 to 14 μcd/m$^2$; blue); from 8 to 50% above natural nighttime brightness (14 to 87 μcd/m$^2$; green); from 50% above natural to the level of light under which the Milky Way is no longer visible (87 to 688 μcd/m$^2$; yellow); from Milky Way loss to estimated cone stimulation (688 to 3000 μcd/m$^2$; red); and very high nighttime light intensities, with no dark adaption for human eyes (>3000 μcd/m$^2$; white).

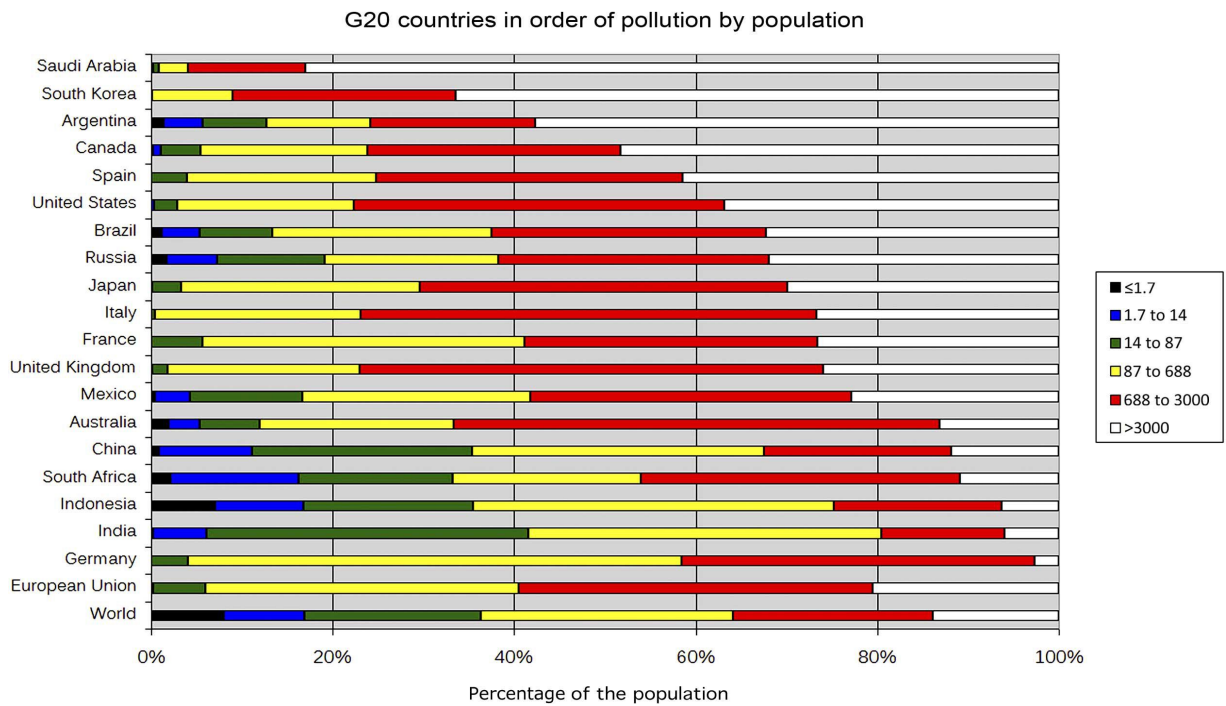

**Fig. 11. G20 countries sorted by population exposed to light pollution.** Countries of the G20 group whose populations live under skies polluted by the specified artificial sky brightness. Color ranges are shown on the right and indicate the pollution level (μcd/m$^2$).

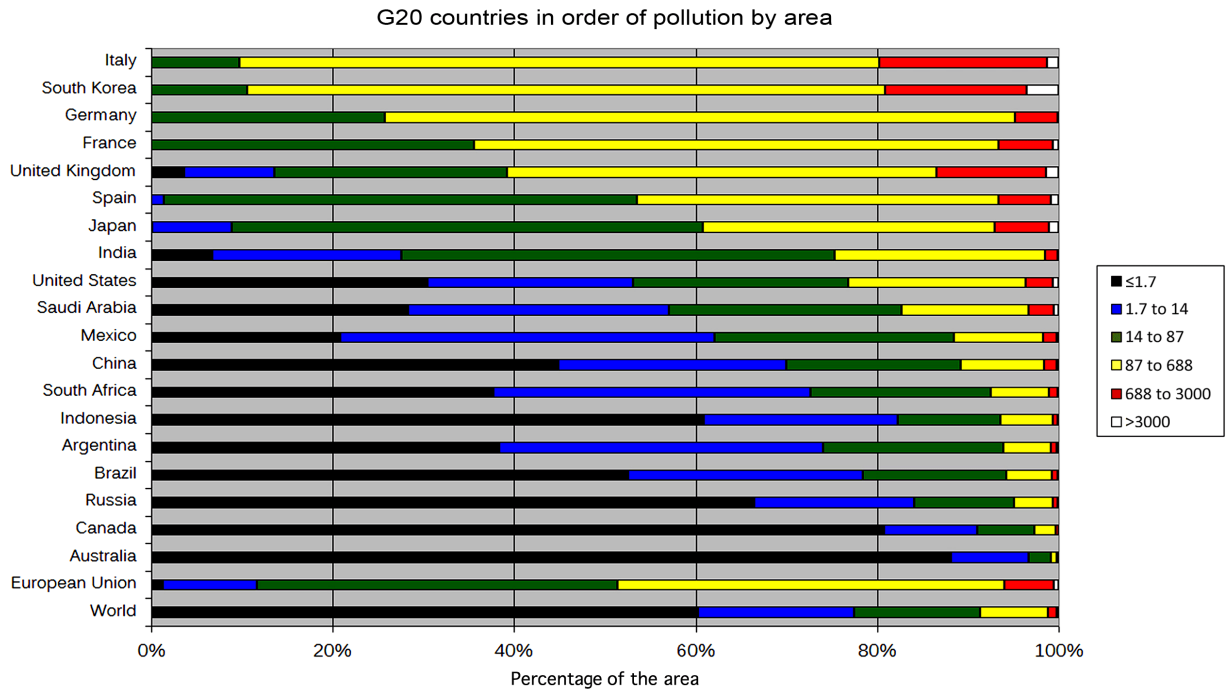

**Fig. 12. G20 countries sorted by polluted area.** Countries of the G20 group whose area is polluted by the specified artificial sky brightness. Countries are ordered using the area of the three most polluted levels (that is, yellow, red, and white). Different orders may be obtained by choosing different pollution levels. Color ranges are shown on the right and indicate the pollution level (μcd/m²).

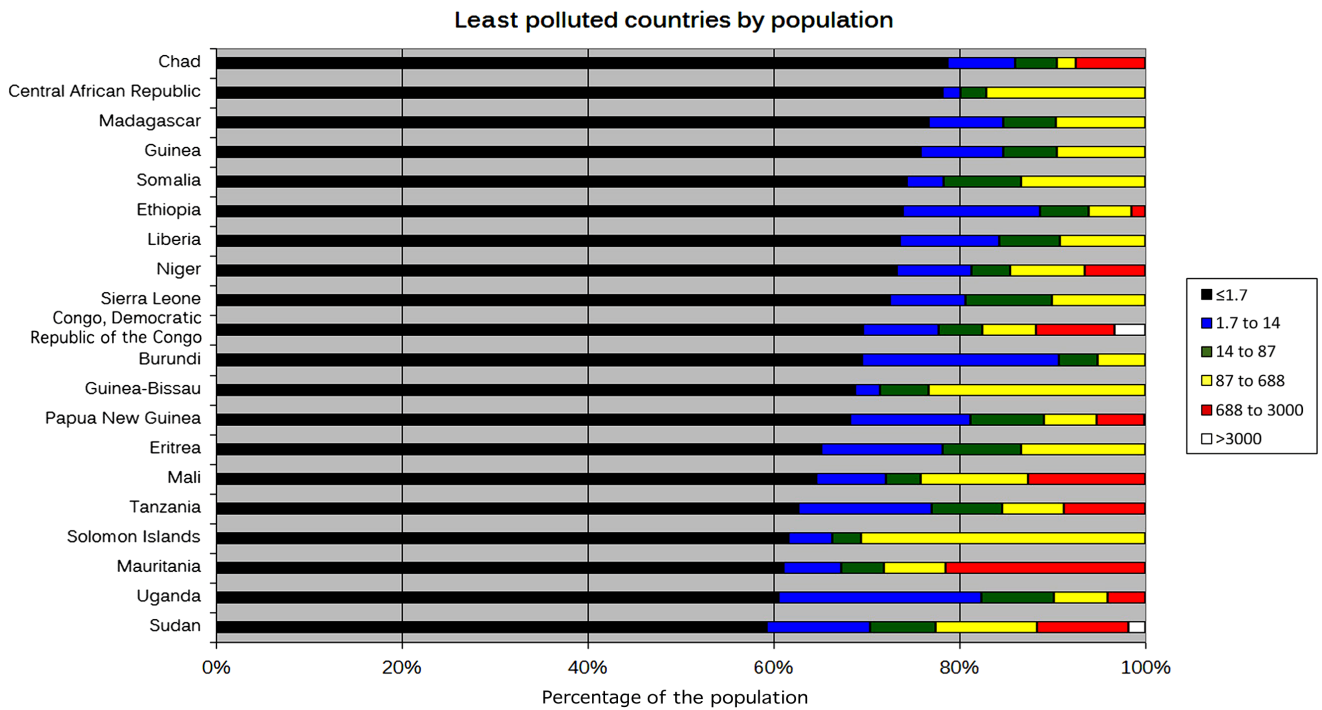

**Fig. 13. The 20 least polluted countries.** Countries whose populations are exposed to the least light pollution. Color ranges are shown on the right and indicate the pollution level (μcd/m²).

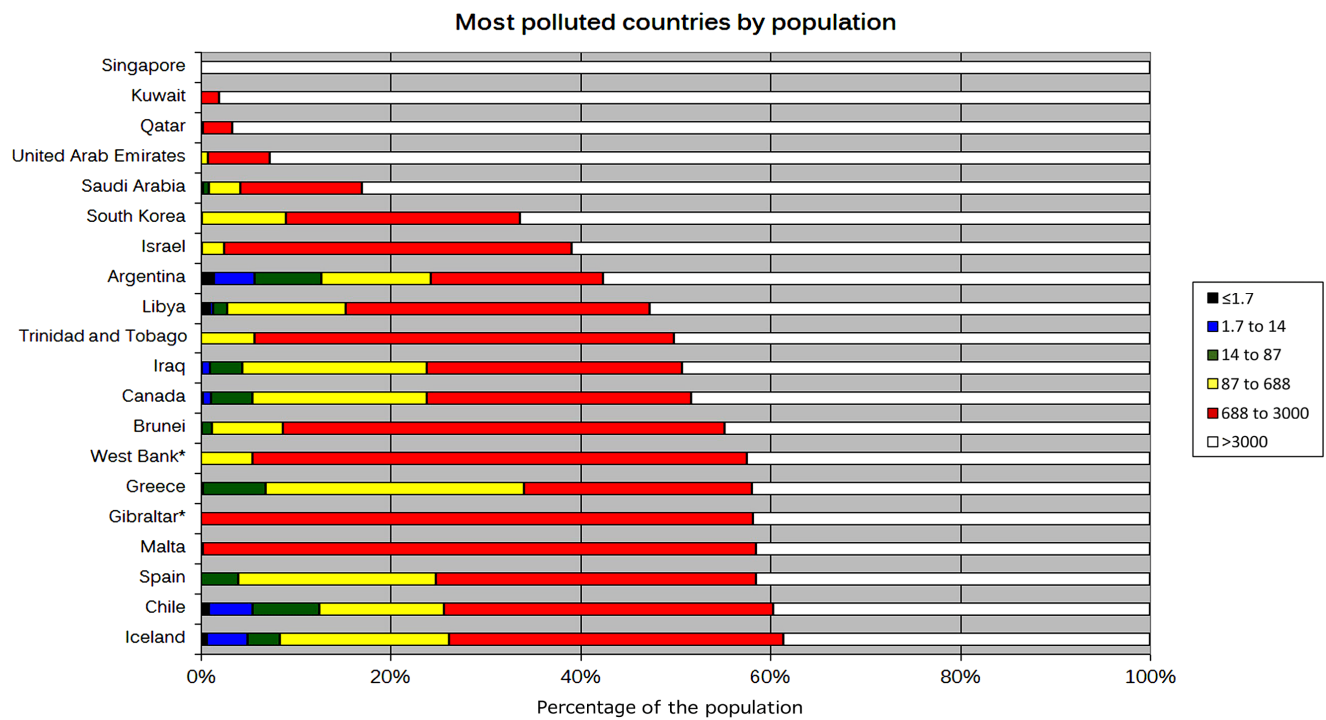

**Fig. 14. The 20 most polluted countries.** Countries whose populations are most exposed to light pollution. Color ranges are shown on the right and indicate the pollution level (μcd/m²).

### Sky brightness data

A collection of night sky brightness observations taken using handheld and vehicle-mounted SQMs was assembled using data provided by both professional researchers and citizen scientists. The data were filtered to remove instances of twilight or moonlight, as well as observations where observers reported problematic conditions (for example, snow or mist). After this process, 20,865 observations remained, with the largest individual contributions from areas near Catalonia (7400), Madrid [see (*40*)] (5355), and Berlin (2371). Globe at Night [see (*41*)] provided a total of 4114 observations, including locations from every continent, with about 20% coming from outside North America or Europe. To reduce the influence of locations with large numbers of observations (for example, 10 or more observations on a single night), we binned the data according to a 30–arcsec grid and assigned an "effective weight" of $(n_e \sqrt{N_T})^{-1}$, where $N_T$ is the total number of nights on which observations were made and $n_e$ is the number of observations taken on the same night.

Multiple observations taken on a single night are not independent. Although they do provide some information about the change in sky radiance over the night, they provide much less information than would an equivalent number of independent observations at widely separated locations. The maximum contribution to the data set from a location with many observations on a single night is therefore set equivalent to a single independent observation. On the other hand, a location where observations are reported on many different nights is likely to include data taken under different atmospheric conditions, days of the week, times, and seasons. These data provide a better description of the typical skyglow at the location than does a single observation, but still not as much information as would an equivalent number of observations from truly independent locations. The uncertainty on the standard deviation (SD) of the mean skyglow radiance should decrease with the square of the number of independent observations, so the weight of the combined observations was increased by a proportional amount. As an example, a single location with five observations taken on four nights contributes the same equivalent weight to the data set as two observations made at two widely separated locations. This weighting procedure led to a total of 10,441 "effective observations."

Observations were adjusted to estimate the artificial sky brightness component by subtracting the natural component computed with a model of V-band natural sky brightness (*42*). The model was customized to the location, date, and time of each observation, and predicted the combined brightness from the Milky Way, zodiacal light, and natural airglow, as measured by an SQM-L instrument aimed at the zenith. The brightness of natural airglow for a given date was predicted on the basis of its relation with solar activity, following the work of Krisciunas *et al.* (*43*).

### Calibration

Maps were produced under three different assumptions for the angular distributions of light intensity emitted upward from cities: one with Lambertian emission (map *A*), one with the highest emission at angles near the horizon (map *B*), and one with a peak intensity at intermediate angles above the horizon (map *C*) (see Fig. 1) (*44*). The predicted zenith total sky luminance (cd/m²) for each observation location is given by $B = SN + (W_a A + W_b B + W_c C)(1 + dh)$, where $N$ is the natural sky

Table 2. Percentages of population and area under the specified artificial sky brightness (μcd/m²).

| Country | Brightness (μcd/m²) | | | | | | | | | | | |
| --- | --- | --- | --- | --- | --- | --- | --- | --- | --- | --- | --- | --- |
| | ≤1.7 | >1.7 | >14 | >87 | >688 | >3000 | ≤1.7 | >1.7 | >14 | >87 | >688 | >3000 |
| | Population (%) | | | | | | Area (%) | | | | | |
| Afghanistan | 39.1 | 60.9 | 37.2 | 26.5 | 11.6 | 0.0 | 79.4 | 20.6 | 4.0 | 0.8 | 0.0 | 0.0 |
| Albania | 0.0 | 100.0 | 98.5 | 75.6 | 37.1 | 10.3 | 0.0 | 100.0 | 87.5 | 23.8 | 1.7 | 0.0 |
| Algeria | 0.3 | 99.7 | 98.5 | 91.9 | 52.1 | 19.2 | 66.4 | 33.6 | 19.3 | 8.7 | 1.1 | 0.2 |
| American Samoa* | 0.0 | 100.0 | 100.0 | 99.7 | 0.0 | 0.0 | 0.0 | 100.0 | 100.0 | 97.0 | 0.0 | 0.0 |
| Andorra | 0.0 | 100.0 | 100.0 | 100.0 | 43.6 | 0.0 | 0.0 | 100.0 | 100.0 | 98.1 | 1.5 | 0.0 |
| Angola | 55.2 | 44.8 | 33.2 | 27.5 | 15.2 | 7.0 | 88.5 | 11.5 | 2.9 | 0.8 | 0.2 | 0.0 |
| Anguilla* | 0.0 | 100.0 | 100.0 | 99.3 | 0.0 | 0.0 | 0.0 | 100.0 | 100.0 | 95.5 | 0.0 | 0.0 |
| Antigua and Barbuda | 0.0 | 100.0 | 100.0 | 100.0 | 58.1 | 0.0 | 0.0 | 100.0 | 100.0 | 98.7 | 13.7 | 0.0 |
| Argentina | 1.4 | 98.6 | 94.3 | 87.3 | 75.8 | 57.7 | 38.4 | 61.6 | 26.0 | 6.1 | 0.9 | 0.2 |
| Armenia | 0.0 | 100.0 | 86.3 | 57.9 | 31.3 | 0.0 | 0.0 | 100.0 | 35.1 | 6.2 | 0.3 | 0.0 |
| Aruba* | 0.0 | 100.0 | 100.0 | 100.0 | 68.3 | 0.0 | 0.0 | 100.0 | 100.0 | 100.0 | 23.9 | 0.0 |
| Australia | 2.0 | 98.0 | 94.7 | 88.0 | 66.7 | 13.1 | 88.1 | 11.9 | 3.4 | 0.9 | 0.2 | 0.0 |
| Austria | 0.0 | 100.0 | 99.9 | 88.7 | 35.7 | 10.3 | 0.0 | 100.0 | 97.8 | 40.0 | 1.8 | 0.1 |
| Azerbaijan | 0.0 | 100.0 | 95.2 | 71.2 | 34.6 | 15.3 | 18.6 | 81.4 | 51.1 | 13.0 | 1.3 | 0.1 |
| Bangladesh | 1.1 | 98.9 | 79.2 | 32.2 | 10.4 | 0.0 | 5.3 | 94.7 | 60.7 | 11.5 | 0.6 | 0.0 |
| Barbados | 0.0 | 100.0 | 100.0 | 100.0 | 6.0 | 0.0 | 0.0 | 100.0 | 100.0 | 99.7 | 1.4 | 0.0 |
| Belarus | 0.0 | 100.0 | 92.7 | 74.9 | 48.5 | 12.0 | 0.0 | 100.0 | 63.2 | 10.3 | 0.9 | 0.1 |
| Belgium | 0.0 | 100.0 | 100.0 | 100.0 | 86.8 | 22.2 | 0.0 | 100.0 | 100.0 | 100.0 | 51.1 | 2.6 |
| Belize | 5.6 | 94.4 | 77.8 | 55.9 | 4.2 | 0.0 | 33.2 | 66.8 | 19.7 | 3.0 | 0.0 | 0.0 |
| Benin | 34.4 | 65.6 | 44.1 | 27.8 | 6.6 | 0.0 | 84.5 | 15.5 | 3.8 | 1.0 | 0.0 | 0.0 |
| Bhutan | 12.3 | 87.7 | 51.0 | 20.7 | 0.0 | 0.0 | 43.6 | 56.4 | 8.6 | 0.4 | 0.0 | 0.0 |
| Bolivia | 13.0 | 87.0 | 72.5 | 63.4 | 48.9 | 12.0 | 77.2 | 22.8 | 5.5 | 1.2 | 0.1 | 0.0 |
| Bosnia and Herzegovina | 0.0 | 100.0 | 98.9 | 79.0 | 26.2 | 0.0 | 0.0 | 100.0 | 88.7 | 26.0 | 0.7 | 0.0 |
| Botswana | 22.7 | 77.3 | 59.3 | 43.0 | 11.5 | 0.0 | 89.6 | 10.4 | 2.2 | 0.4 | 0.0 | 0.0 |
| Brazil | 1.2 | 98.8 | 94.7 | 86.7 | 62.5 | 32.3 | 52.6 | 47.4 | 21.6 | 5.7 | 0.7 | 0.1 |
| British Virgin Islands* | 0.0 | 100.0 | 100.0 | 100.0 | 0.0 | 0.0 | 0.0 | 100.0 | 100.0 | 100.0 | 0.0 | 0.0 |
| Brunei | 0.0 | 100.0 | 99.9 | 98.8 | 91.3 | 44.9 | 0.0 | 100.0 | 76.4 | 44.8 | 12.5 | 1.6 |
| Bulgaria | 0.0 | 100.0 | 99.2 | 77.3 | 37.0 | 4.9 | 0.0 | 100.0 | 94.6 | 19.5 | 0.9 | 0.0 |
| Burkina Faso | 59.2 | 40.8 | 23.4 | 16.9 | 10.9 | 0.0 | 84.5 | 15.5 | 2.7 | 0.5 | 0.1 | 0.0 |
| Burundi | 69.5 | 30.5 | 9.3 | 5.2 | 0.0 | 0.0 | 82.8 | 17.2 | 1.9 | 0.3 | 0.0 | 0.0 |
| Cambodia | 30.0 | 70.0 | 30.9 | 17.2 | 9.8 | 0.0 | 72.4 | 27.6 | 4.4 | 0.8 | 0.1 | 0.0 |
| Cameroon | 44.6 | 55.4 | 42.1 | 29.3 | 19.9 | 0.0 | 88.0 | 12.0 | 3.0 | 0.5 | 0.1 | 0.0 |
| Canada | 0.2 | 99.8 | 98.9 | 94.6 | 76.2 | 48.3 | 80.8 | 19.2 | 9.0 | 2.7 | 0.3 | 0.1 |
| Cape Verde | 1.4 | 98.6 | 80.8 | 42.7 | 30.2 | 0.0 | 24.6 | 75.4 | 25.9 | 4.0 | 0.5 | 0.0 |
| Cayman Islands* | 0.0 | 100.0 | 100.0 | 99.9 | 67.6 | 0.0 | 0.0 | 100.0 | 100.0 | 96.9 | 14.0 | 0.0 |
|  | | | | | | | | | | | | |

| Country | Brightness (μcd/m²) | | | | | | | | | | | |
|---|---|---|---|---|---|---|---|---|---|---|---|---|
| | ≤1.7 | >1.7 | >14 | >87 | >688 | >3000 | ≤1.7 | >1.7 | >14 | >87 | >688 | >3000 |
| | Population (%) | | | | | | Area (%) | | | | | |
| Central African Republic | 78.2 | 21.8 | 19.9 | 17.1 | 0.0 | 0.0 | 99.7 | 0.3 | 0.1 | 0.0 | 0.0 | 0.0 |
| Chad | 78.7 | 21.3 | 14.0 | 9.5 | 7.5 | 0.0 | 98.2 | 1.8 | 0.5 | 0.1 | 0.0 | 0.0 |
| Chile | 0.8 | 99.2 | 94.5 | 87.5 | 74.4 | 39.7 | 51.4 | 48.6 | 18.4 | 5.1 | 0.7 | 0.1 |
| China | 0.9 | 99.1 | 88.9 | 64.6 | 32.5 | 11.9 | 44.8 | 55.2 | 30.0 | 10.8 | 1.6 | 0.2 |
| Christmas Island* | 0.0 | 100.0 | 99.9 | 74.3 | 0.0 | 0.0 | 0.0 | 100.0 | 49.1 | 6.9 | 0.0 | 0.0 |
| Cocos Islands* | 100.0 | 0.0 | 0.0 | 0.0 | 0.0 | 0.0 | 100.0 | 0.0 | 0.0 | 0.0 | 0.0 | 0.0 |
| Colombia | 3.0 | 97.0 | 89.3 | 75.0 | 54.6 | 18.7 | 55.8 | 44.2 | 22.1 | 5.2 | 0.5 | 0.0 |
| Comoros | 38.2 | 61.8 | 38.2 | 15.6 | 0.0 | 0.0 | 64.5 | 35.5 | 6.2 | 0.8 | 0.0 | 0.0 |
| Congo | 26.2 | 73.8 | 64.9 | 57.6 | 49.0 | 4.7 | 87.6 | 12.4 | 3.6 | 1.1 | 0.2 | 0.0 |
| Congo, Democratic Republic of the Congo | 69.6 | 30.4 | 22.2 | 17.5 | 11.8 | 3.3 | 95.8 | 4.2 | 1.0 | 0.2 | 0.0 | 0.0 |
| Cook Islands* | 0.0 | 100.0 | 85.4 | 0.0 | 0.0 | 0.0 | 0.0 | 100.0 | 34.7 | 0.0 | 0.0 | 0.0 |
| Costa Rica | 0.0 | 100.0 | 97.3 | 81.8 | 52.9 | 2.4 | 0.4 | 99.6 | 70.4 | 18.3 | 1.6 | 0.0 |
| Cote d'Ivoire | 17.8 | 82.2 | 54.2 | 33.6 | 19.6 | 3.5 | 50.4 | 49.6 | 11.1 | 1.5 | 0.2 | 0.0 |
| Croatia | 0.0 | 100.0 | 100.0 | 95.2 | 50.5 | 21.2 | 0.0 | 100.0 | 98.9 | 62.2 | 4.0 | 0.4 |
| Cuba | 0.6 | 99.4 | 90.0 | 66.2 | 39.5 | 1.9 | 6.7 | 93.3 | 52.3 | 10.9 | 0.9 | 0.0 |
| Cyprus | 0.0 | 100.0 | 99.9 | 98.1 | 71.4 | 10.5 | 0.0 | 100.0 | 99.1 | 72.2 | 8.4 | 0.2 |
| Czech Republic | 0.0 | 100.0 | 100.0 | 97.5 | 42.8 | 7.3 | 0.0 | 100.0 | 100.0 | 82.3 | 3.6 | 0.2 |
| Denmark | 0.0 | 100.0 | 99.9 | 89.3 | 38.5 | 7.3 | 0.0 | 100.0 | 99.2 | 47.9 | 2.9 | 0.1 |
| Djibouti | 48.9 | 51.1 | 39.7 | 0.8 | 0.0 | 0.0 | 82.3 | 17.7 | 2.8 | 0.1 | 0.0 | 0.0 |
| Dominica | 0.0 | 100.0 | 94.5 | 43.7 | 0.0 | 0.0 | 0.0 | 100.0 | 44.4 | 5.7 | 0.0 | 0.0 |
| Dominican Republic | 0.0 | 100.0 | 98.1 | 82.3 | 57.6 | 22.6 | 0.2 | 99.8 | 79.8 | 23.0 | 2.8 | 0.3 |
| Ecuador | 0.4 | 99.6 | 95.3 | 78.3 | 50.0 | 17.7 | 17.2 | 82.8 | 53.4 | 14.9 | 1.7 | 0.2 |
| Egypt | 0.0 | 100.0 | 99.9 | 99.8 | 97.5 | 37.1 | 52.4 | 47.6 | 26.5 | 12.6 | 4.9 | 0.5 |
| El Salvador | 0.0 | 100.0 | 94.9 | 67.3 | 24.2 | 0.0 | 0.0 | 100.0 | 80.2 | 18.7 | 1.1 | 0.0 |
| Equatorial Guinea | 17.2 | 82.8 | 54.2 | 39.1 | 20.0 | 0.0 | 20.4 | 79.6 | 22.3 | 3.6 | 0.3 | 0.0 |
| Eritrea | 65.1 | 34.9 | 21.9 | 13.3 | 0.0 | 0.0 | 95.5 | 4.5 | 0.7 | 0.1 | 0.0 | 0.0 |
| Estonia | 0.2 | 99.8 | 97.4 | 83.9 | 60.5 | 31.8 | 2.7 | 97.3 | 75.1 | 19.7 | 2.3 | 0.3 |
| Ethiopia | 73.9 | 26.1 | 11.3 | 6.1 | 1.5 | 0.0 | 93.4 | 6.6 | 1.2 | 0.2 | 0.0 | 0.0 |
| Falkland Islands* | 17.6 | 82.4 | 80.2 | 67.8 | 0.0 | 0.0 | 88.7 | 11.3 | 1.8 | 0.2 | 0.0 | 0.0 |
| Faroe Islands* | 0.0 | 100.0 | 96.5 | 74.5 | 12.6 | 0.0 | 0.1 | 99.9 | 84.3 | 21.5 | 0.3 | 0.0 |
| Fiji | 27.4 | 72.6 | 49.1 | 29.9 | 0.0 | 0.0 | 67.7 | 32.3 | 7.5 | 1.3 | 0.0 | 0.0 |
| Finland | 0.0 | 100.0 | 99.7 | 95.7 | 68.0 | 33.7 | 2.8 | 97.2 | 73.9 | 27.6 | 2.9 | 0.4 |
| France | 0.0 | 100.0 | 100.0 | 94.3 | 58.9 | 26.6 | 0.0 | 100.0 | 100.0 | 64.5 | 6.7 | 0.7 |
| French Guiana* | 7.8 | 92.2 | 88.4 | 77.9 | 36.7 | 0.0 | 88.3 | 11.7 | 2.9 | 0.7 | 0.1 | 0.0 |
| French Polynesia* | 0.0 | 100.0 | 98.7 | 80.8 | 4.9 | 0.0 | 0.0 | 100.0 | 59.6 | 10.9 | 0.2 | 0.0 |
| Gabon | 22.0 | 78.0 | 69.4 | 56.9 | 38.7 | 0.2 | 77.5 | 22.5 | 7.9 | 2.0 | 0.3 | 0.0 |
| continued on next page | | | | | | | | | | | | |

| Country | Brightness (μcd/m²) | | | | | | | | | | | |
|---|---|---|---|---|---|---|---|---|---|---|---|---|
| | ≤1.7 | >1.7 | >14 | >87 | >688 | >3000 | ≤1.7 | >1.7 | >14 | >87 | >688 | >3000 |
| | Population (%) | | | | | | Area (%) | | | | | |
| Gaza Strip* | 0.0 | 100.0 | 100.0 | 100.0 | 95.1 | 0.0 | 0.0 | 100.0 | 100.0 | 100.0 | 68.1 | 0.0 |
| Georgia | 0.0 | 100.0 | 92.2 | 63.6 | 36.7 | 4.5 | 0.0 | 100.0 | 49.8 | 10.8 | 0.8 | 0.0 |
| Germany | 0.0 | 100.0 | 100.0 | 96.0 | 41.6 | 2.7 | 0.0 | 100.0 | 100.0 | 74.2 | 4.8 | 0.1 |
| Ghana | 15.5 | 84.5 | 60.6 | 37.1 | 21.4 | 0.4 | 51.5 | 48.5 | 16.3 | 3.4 | 0.5 | 0.0 |
| Gibraltar* | 0.0 | 100.0 | 100.0 | 100.0 | 100.0 | 41.9 | 0.0 | 100.0 | 100.0 | 100.0 | 100.0 | 28.6 |
| Greece | 0.0 | 100.0 | 99.8 | 93.2 | 66.0 | 41.9 | 0.0 | 100.0 | 96.3 | 40.1 | 3.0 | 0.5 |
| Greenland* | 13.3 | 86.7 | 86.6 | 78.1 | 3.9 | 0.0 | 99.9 | 0.1 | 0.0 | 0.0 | 0.0 | 0.0 |
| Grenada | 0.0 | 100.0 | 100.0 | 76.3 | 0.0 | 0.0 | 0.0 | 100.0 | 100.0 | 39.8 | 0.0 | 0.0 |
| Guadeloupe* | 0.0 | 100.0 | 100.0 | 99.7 | 60.2 | 1.7 | 0.0 | 100.0 | 100.0 | 87.5 | 11.3 | 0.1 |
| Guam* | 0.0 | 100.0 | 100.0 | 100.0 | 82.9 | 0.0 | 0.0 | 100.0 | 100.0 | 95.8 | 27.4 | 0.0 |
| Guatemala | 2.2 | 97.8 | 84.8 | 46.3 | 20.2 | 0.5 | 25.4 | 74.6 | 38.9 | 7.2 | 0.6 | 0.0 |
| Guernsey* | 0.0 | 100.0 | 100.0 | 91.1 | 0.0 | 0.0 | 0.0 | 100.0 | 100.0 | 68.7 | 0.0 | 0.0 |
| Guinea | 75.8 | 24.2 | 15.3 | 9.6 | 0.0 | 0.0 | 95.1 | 4.9 | 0.8 | 0.2 | 0.0 | 0.0 |
| Guinea-Bissau | 68.7 | 31.3 | 28.5 | 23.3 | 0.0 | 0.0 | 97.5 | 2.5 | 0.6 | 0.1 | 0.0 | 0.0 |
| Guyana | 26.3 | 73.7 | 55.5 | 43.5 | 5.8 | 0.0 | 94.3 | 5.7 | 1.2 | 0.2 | 0.0 | 0.0 |
| Haiti | 28.1 | 71.9 | 41.1 | 30.7 | 19.2 | 0.0 | 43.5 | 56.5 | 12.6 | 2.5 | 0.3 | 0.0 |
| Honduras | 2.1 | 97.9 | 78.5 | 51.2 | 33.2 | 2.2 | 28.9 | 71.1 | 34.2 | 6.0 | 0.5 | 0.0 |
| Hungary | 0.0 | 100.0 | 100.0 | 86.0 | 38.5 | 9.8 | 0.0 | 100.0 | 100.0 | 41.1 | 2.1 | 0.2 |
| Iceland | 0.7 | 99.3 | 95.0 | 91.6 | 73.8 | 38.7 | 36.0 | 64.0 | 24.4 | 7.6 | 0.6 | 0.1 |
| India | 0.2 | 99.8 | 93.9 | 58.5 | 19.5 | 5.9 | 6.8 | 93.2 | 72.5 | 24.7 | 1.5 | 0.1 |
| Indonesia | 7.0 | 93.0 | 83.2 | 64.5 | 24.8 | 6.3 | 60.9 | 39.1 | 17.7 | 6.4 | 0.6 | 0.1 |
| Iran | 0.2 | 99.8 | 97.9 | 88.5 | 64.0 | 17.1 | 17.0 | 83.0 | 55.5 | 18.1 | 2.6 | 0.3 |
| Iraq | 0.1 | 99.9 | 99.1 | 95.6 | 76.2 | 49.4 | 28.7 | 71.3 | 53.7 | 33.2 | 9.0 | 2.5 |
| Ireland | 0.0 | 100.0 | 99.6 | 83.9 | 45.2 | 18.5 | 0.0 | 100.0 | 94.6 | 39.4 | 2.0 | 0.3 |
| Isle of Man* | 0.0 | 100.0 | 99.8 | 77.4 | 42.7 | 0.0 | 0.0 | 100.0 | 97.8 | 31.5 | 2.8 | 0.0 |
| Israel | 0.0 | 100.0 | 100.0 | 99.9 | 97.6 | 61.0 | 0.0 | 100.0 | 98.2 | 76.3 | 41.9 | 8.1 |
| Italy | 0.0 | 100.0 | 100.0 | 99.6 | 76.9 | 26.7 | 0.0 | 100.0 | 100.0 | 90.3 | 19.7 | 1.3 |
| Jamaica | 0.0 | 100.0 | 100.0 | 86.0 | 47.2 | 4.1 | 0.0 | 100.0 | 100.0 | 41.7 | 2.9 | 0.1 |
| Japan | 0.0 | 100.0 | 99.9 | 96.7 | 70.4 | 29.9 | 0.1 | 99.9 | 91.1 | 39.2 | 7.1 | 1.0 |
| Jersey* | 0.0 | 100.0 | 100.0 | 98.9 | 35.8 | 0.0 | 0.0 | 100.0 | 100.0 | 91.5 | 6.5 | 0.0 |
| Jordan | 0.1 | 99.9 | 99.7 | 98.8 | 80.5 | 24.7 | 13.6 | 86.4 | 52.0 | 22.2 | 4.2 | 0.2 |
| Kazakhstan | 7.7 | 92.3 | 80.5 | 66.0 | 45.0 | 12.3 | 60.9 | 39.1 | 11.3 | 2.9 | 0.5 | 0.1 |
| Kenya | 34.9 | 65.1 | 31.6 | 18.3 | 9.1 | 0.0 | 85.2 | 14.8 | 3.3 | 0.7 | 0.1 | 0.0 |
| Kuwait | 0.0 | 100.0 | 100.0 | 100.0 | 100.0 | 98.1 | 0.0 | 100.0 | 100.0 | 92.0 | 50.9 | 11.5 |
| Kyrgyzstan | 1.8 | 98.2 | 88.8 | 60.5 | 18.6 | 0.0 | 35.9 | 64.1 | 19.3 | 3.6 | 0.1 | 0.0 |
| Laos | 41.0 | 59.0 | 35.7 | 20.1 | 8.3 | 0.0 | 73.7 | 26.3 | 5.0 | 1.0 | 0.1 | 0.0 |
| continued on next page | | | | | | | | | | | | |

| Country | Brightness (μcd/m²) | | | | | | | | | | | |
|---|---|---|---|---|---|---|---|---|---|---|---|---|
| | ≤1.7 | >1.7 | >14 | >87 | >688 | >3000 | ≤1.7 | >1.7 | >14 | >87 | >688 | >3000 |
| | Population (%) | | | | | | Area (%) | | | | | |
| Latvia | 0.0 | 100.0 | 89.2 | 71.9 | 46.8 | 21.2 | 0.0 | 100.0 | 48.3 | 9.2 | 1.0 | 0.1 |
| Lebanon | 0.0 | 100.0 | 100.0 | 99.6 | 63.0 | 28.4 | 0.0 | 100.0 | 100.0 | 88.5 | 18.3 | 1.7 |
| Lesotho | 18.1 | 81.9 | 45.9 | 21.8 | 0.0 | 0.0 | 45.2 | 54.8 | 9.2 | 1.0 | 0.0 | 0.0 |
| Liberia | 73.6 | 26.4 | 15.7 | 9.2 | 0.0 | 0.0 | 95.1 | 4.9 | 0.9 | 0.1 | 0.0 | 0.0 |
| Libya | 0.9 | 99.1 | 98.7 | 97.3 | 84.7 | 52.7 | 71.1 | 28.9 | 12.4 | 4.1 | 0.7 | 0.1 |
| Liechtenstein | 0.0 | 100.0 | 100.0 | 100.0 | 0.0 | 0.0 | 0.0 | 100.0 | 100.0 | 93.7 | 0.0 | 0.0 |
| Lithuania | 0.0 | 100.0 | 93.0 | 66.7 | 41.9 | 7.8 | 0.0 | 100.0 | 72.0 | 11.7 | 1.0 | 0.0 |
| Luxembourg | 0.0 | 100.0 | 100.0 | 100.0 | 60.1 | 6.9 | 0.0 | 100.0 | 100.0 | 100.0 | 18.9 | 0.2 |
| Macedonia | 0.0 | 100.0 | 100.0 | 84.2 | 42.4 | 10.8 | 0.0 | 100.0 | 100.0 | 23.4 | 1.4 | 0.1 |
| Madagascar | 76.7 | 23.3 | 15.3 | 9.6 | 0.0 | 0.0 | 97.4 | 2.6 | 0.5 | 0.1 | 0.0 | 0.0 |
| Malawi | 29.7 | 70.3 | 27.9 | 13.4 | 0.7 | 0.0 | 62.3 | 37.7 | 6.5 | 1.1 | 0.0 | 0.0 |
| Malaysia | 1.7 | 98.3 | 94.2 | 88.9 | 67.9 | 34.6 | 32.3 | 67.7 | 40.7 | 19.8 | 3.7 | 0.6 |
| Mali | 64.6 | 35.4 | 27.9 | 24.1 | 12.6 | 0.0 | 97.0 | 3.0 | 0.6 | 0.1 | 0.0 | 0.0 |
| Malta | 0.0 | 100.0 | 100.0 | 100.0 | 99.8 | 41.5 | 0.0 | 100.0 | 100.0 | 100.0 | 88.5 | 16.7 |
| Martinique* | 0.0 | 100.0 | 100.0 | 100.0 | 65.5 | 7.2 | 0.0 | 100.0 | 100.0 | 98.8 | 25.2 | 0.3 |
| Mauritania | 61.1 | 38.9 | 32.7 | 28.2 | 21.4 | 0.0 | 98.6 | 1.4 | 0.3 | 0.1 | 0.0 | 0.0 |
| Mauritius | 0.0 | 100.0 | 100.0 | 95.1 | 29.9 | 0.0 | 0.0 | 100.0 | 100.0 | 70.9 | 4.3 | 0.0 |
| Mayotte* | 0.0 | 100.0 | 100.0 | 69.5 | 0.0 | 0.0 | 0.0 | 100.0 | 100.0 | 37.6 | 0.0 | 0.0 |
| Mexico | 0.5 | 99.5 | 95.8 | 83.3 | 58.3 | 22.8 | 20.8 | 79.2 | 37.9 | 11.6 | 1.7 | 0.2 |
| Midway Islands* | 100.0 | 0.0 | 0.0 | 0.0 | 0.0 | 0.0 | 100.0 | 0.0 | 0.0 | 0.0 | 0.0 | 0.0 |
| Moldova | 0.0 | 100.0 | 89.9 | 43.8 | 18.6 | 0.0 | 0.0 | 100.0 | 68.7 | 8.2 | 0.5 | 0.0 |
| Mongolia | 31.4 | 68.6 | 63.2 | 51.6 | 36.7 | 0.0 | 95.6 | 4.4 | 0.8 | 0.1 | 0.0 | 0.0 |
| Montenegro | 0.0 | 100.0 | 97.2 | 80.6 | 44.7 | 11.7 | 0.0 | 100.0 | 81.0 | 24.9 | 1.5 | 0.1 |
| Montserrat* | 0.0 | 100.0 | 97.8 | 0.0 | 0.0 | 0.0 | 0.0 | 100.0 | 17.9 | 0.0 | 0.0 | 0.0 |
| Morocco | 0.9 | 99.1 | 91.0 | 67.5 | 49.3 | 29.3 | 26.6 | 73.4 | 39.8 | 11.3 | 1.4 | 0.2 |
| Mozambique | 55.8 | 44.2 | 28.3 | 20.3 | 11.9 | 2.9 | 87.2 | 12.8 | 2.9 | 0.6 | 0.1 | 0.0 |
| Myanmar | 26.2 | 73.8 | 39.9 | 21.5 | 9.5 | 0.6 | 70.3 | 29.7 | 5.6 | 1.0 | 0.1 | 0.0 |
| Namibia | 31.3 | 68.7 | 50.1 | 37.1 | 17.4 | 0.0 | 92.3 | 7.7 | 1.5 | 0.2 | 0.0 | 0.0 |
| Nauru | 0.0 | 100.0 | 100.0 | 100.0 | 0.0 | 0.0 | 0.0 | 100.0 | 100.0 | 100.0 | 0.0 | 0.0 |
| Nepal | 21.9 | 78.1 | 45.7 | 18.1 | 0.0 | 0.0 | 60.5 | 39.5 | 10.2 | 1.2 | 0.0 | 0.0 |
| Netherlands | 0.0 | 100.0 | 100.0 | 100.0 | 81.2 | 26.2 | 0.0 | 100.0 | 100.0 | 99.4 | 42.5 | 7.2 |
| Netherlands Antilles* | 0.0 | 100.0 | 100.0 | 99.2 | 93.5 | 0.7 | 0.0 | 100.0 | 100.0 | 79.6 | 30.0 | 1.7 |
| New Caledonia* | 7.6 | 92.4 | 79.2 | 67.2 | 48.4 | 0.0 | 38.9 | 61.1 | 16.3 | 3.9 | 0.6 | 0.0 |
| New Zealand | 2.8 | 97.2 | 91.4 | 83.4 | 56.4 | 4.8 | 53.1 | 46.9 | 15.1 | 3.5 | 0.5 | 0.0 |
| Nicaragua | 16.9 | 83.1 | 65.5 | 50.7 | 24.3 | 0.0 | 64.7 | 35.3 | 11.8 | 2.5 | 0.2 | 0.0 |
| Niger | 73.2 | 26.8 | 18.7 | 14.6 | 6.5 | 0.0 | 97.3 | 2.7 | 0.6 | 0.1 | 0.0 | 0.0 |
| continued on next page | | | | | | | | | | | | |

| Country | Brightness (μcd/m²) | | | | | | | | | | | |
|---|---|---|---|---|---|---|---|---|---|---|---|---|
| | ≤1.7 | >1.7 | >14 | >87 | >688 | >3000 | ≤1.7 | >1.7 | >14 | >87 | >688 | >3000 |
| | Population (%) | | | | | | Area (%) | | | | | |
| Nigeria | 25.4 | 74.6 | 52.6 | 35.6 | 13.8 | 0.5 | 59.7 | 40.3 | 15.8 | 7.0 | 1.4 | 0.2 |
| Niue* | 97.3 | 2.7 | 0.0 | 0.0 | 0.0 | 0.0 | 99.5 | 0.5 | 0.0 | 0.0 | 0.0 | 0.0 |
| Norfolk Island* | 100.0 | 0.0 | 0.0 | 0.0 | 0.0 | 0.0 | 100.0 | 0.0 | 0.0 | 0.0 | 0.0 | 0.0 |
| North Korea | 27.3 | 72.7 | 42.4 | 15.4 | 1.1 | 0.0 | 61.9 | 38.1 | 10.0 | 1.8 | 0.0 | 0.0 |
| Northern Mariana Islands* | 0.0 | 100.0 | 100.0 | 100.0 | 0.0 | 0.0 | 0.0 | 100.0 | 100.0 | 100.0 | 0.0 | 0.0 |
| Norway | 0.0 | 100.0 | 99.7 | 95.5 | 69.1 | 28.7 | 1.4 | 98.6 | 66.1 | 22.1 | 1.8 | 0.2 |
| Oman | 0.1 | 99.9 | 98.9 | 96.1 | 80.5 | 34.1 | 15.1 | 84.9 | 54.3 | 23.2 | 3.2 | 0.5 |
| Pakistan | 3.4 | 96.6 | 89.8 | 58.6 | 19.7 | 6.6 | 45.9 | 54.1 | 32.2 | 9.4 | 0.5 | 0.1 |
| Panama | 6.3 | 93.7 | 86.0 | 73.0 | 54.1 | 17.2 | 42.5 | 57.5 | 27.0 | 7.4 | 1.2 | 0.1 |
| Papua New Guinea | 68.3 | 31.7 | 18.8 | 10.9 | 5.3 | 0.1 | 89.4 | 10.6 | 2.5 | 0.6 | 0.1 | 0.0 |
| Paraguay | 1.6 | 98.4 | 86.4 | 70.9 | 55.0 | 34.7 | 57.2 | 42.8 | 18.2 | 3.9 | 0.6 | 0.1 |
| Peru | 8.7 | 91.3 | 73.2 | 63.4 | 48.3 | 16.4 | 62.4 | 37.6 | 8.7 | 1.7 | 0.2 | 0.0 |
| Philippines | 8.7 | 91.3 | 67.2 | 47.6 | 26.5 | 7.2 | 35.6 | 64.4 | 20.9 | 6.1 | 0.7 | 0.1 |
| Poland | 0.0 | 100.0 | 100.0 | 93.9 | 50.1 | 13.9 | 0.0 | 100.0 | 99.7 | 67.5 | 4.5 | 0.3 |
| Portugal | 0.0 | 100.0 | 100.0 | 98.3 | 76.7 | 35.4 | 0.0 | 100.0 | 100.0 | 71.2 | 12.8 | 1.3 |
| Puerto Rico* | 0.0 | 100.0 | 100.0 | 100.0 | 71.0 | 25.5 | 0.0 | 100.0 | 100.0 | 99.9 | 28.6 | 3.4 |
| Qatar | 0.0 | 100.0 | 100.0 | 100.0 | 99.8 | 96.7 | 0.0 | 100.0 | 100.0 | 97.0 | 54.6 | 16.3 |
| Reunion* | 0.0 | 100.0 | 100.0 | 95.2 | 43.8 | 0.0 | 0.0 | 100.0 | 99.7 | 51.0 | 3.7 | 0.0 |
| Romania | 0.0 | 100.0 | 99.3 | 65.7 | 32.3 | 10.5 | 0.0 | 100.0 | 92.5 | 24.3 | 1.3 | 0.1 |
| Russia | 1.7 | 98.3 | 92.7 | 80.9 | 61.8 | 32.0 | 66.5 | 33.5 | 16.0 | 4.9 | 0.6 | 0.1 |
| Rwanda | 9.9 | 90.1 | 28.0 | 10.2 | 0.5 | 0.0 | 25.0 | 75.0 | 13.4 | 1.9 | 0.0 | 0.0 |
| Samoa | 31.7 | 68.3 | 62.1 | 35.6 | 0.0 | 0.0 | 79.6 | 20.4 | 4.3 | 0.8 | 0.0 | 0.0 |
| San Marino | 0.0 | 100.0 | 100.0 | 100.0 | 100.0 | 0.0 | 0.0 | 100.0 | 100.0 | 100.0 | 100.0 | 0.0 |
| Sao Tome and Principe | 5.6 | 94.4 | 80.5 | 41.0 | 0.0 | 0.0 | 25.5 | 74.5 | 26.1 | 5.2 | 0.0 | 0.0 |
| Saudi Arabia | 0.0 | 100.0 | 99.8 | 99.1 | 95.9 | 83.0 | 28.3 | 71.7 | 43.0 | 17.3 | 3.3 | 0.6 |
| Senegal | 33.5 | 66.5 | 45.5 | 34.0 | 15.5 | 0.0 | 81.9 | 18.1 | 4.3 | 0.9 | 0.1 | 0.0 |
| Serbia | 0.0 | 100.0 | 100.0 | 92.4 | 39.4 | 13.6 | 0.0 | 100.0 | 100.0 | 57.9 | 3.1 | 0.2 |
| Seychelles | 0.0 | 100.0 | 100.0 | 76.5 | 4.6 | 0.0 | 0.0 | 100.0 | 99.6 | 55.6 | 0.4 | 0.0 |
| Sierra Leone | 72.5 | 27.5 | 19.3 | 10.1 | 0.0 | 0.0 | 91.1 | 8.9 | 1.3 | 0.1 | 0.0 | 0.0 |
| Singapore | 0.0 | 100.0 | 100.0 | 100.0 | 100.0 | 100.0 | 0.0 | 100.0 | 100.0 | 100.0 | 100.0 | 100.0 |
| Slovakia | 0.0 | 100.0 | 99.8 | 82.3 | 23.8 | 4.8 | 0.0 | 100.0 | 98.6 | 46.8 | 1.8 | 0.2 |
| Slovenia | 0.0 | 100.0 | 100.0 | 95.9 | 27.0 | 0.4 | 0.0 | 100.0 | 100.0 | 68.4 | 1.9 | 0.0 |
| Solomon Islands | 61.6 | 38.4 | 33.7 | 30.6 | 0.0 | 0.0 | 94.3 | 5.7 | 1.2 | 0.2 | 0.0 | 0.0 |
| Somalia | 74.4 | 25.6 | 21.7 | 13.3 | 0.0 | 0.0 | 98.8 | 1.2 | 0.2 | 0.0 | 0.0 | 0.0 |
| South Africa | 2.2 | 97.8 | 83.7 | 66.8 | 46.1 | 10.9 | 37.7 | 62.3 | 27.4 | 7.5 | 1.1 | 0.1 |
| South Korea | 0.0 | 100.0 | 100.0 | 99.8 | 91.0 | 66.4 | 0.0 | 100.0 | 100.0 | 89.4 | 19.1 | 3.5 |
| continued on next page | | | | | | | | | | | | |

| Country | Brightness (μcd/m²) | | | | | | | | | | | |
|---|---|---|---|---|---|---|---|---|---|---|---|---|
| | ≤1.7 | >1.7 | >14 | >87 | >688 | >3000 | ≤1.7 | >1.7 | >14 | >87 | >688 | >3000 |
| | Population (%) | | | | | | Area (%) | | | | | |
| Spain | 0.0 | 100.0 | 100.0 | 96.1 | 75.2 | 41.5 | 0.0 | 100.0 | 98.6 | 46.5 | 6.7 | 0.9 |
| Sri Lanka | 0.7 | 99.3 | 76.4 | 29.8 | 2.5 | 0.0 | 11.4 | 88.6 | 33.9 | 5.4 | 0.2 | 0.0 |
| St. Helena* | 16.1 | 83.9 | 0.0 | 0.0 | 0.0 | 0.0 | 51.4 | 48.6 | 0.9 | 0.0 | 0.0 | 0.0 |
| St. Kitts and Nevis | 0.0 | 100.0 | 100.0 | 99.5 | 59.7 | 0.0 | 0.0 | 100.0 | 100.0 | 56.9 | 5.0 | 0.0 |
| St. Lucia | 0.0 | 100.0 | 100.0 | 96.7 | 44.2 | 0.0 | 0.0 | 100.0 | 100.0 | 63.1 | 1.8 | 0.0 |
| St. Pierre and Miquelon* | 3.7 | 96.3 | 48.1 | 48.1 | 0.0 | 0.0 | 26.2 | 73.8 | 4.3 | 0.4 | 0.0 | 0.0 |
| Sudan | 59.3 | 40.7 | 29.6 | 22.5 | 11.6 | 1.9 | 93.9 | 6.1 | 1.6 | 0.4 | 0.1 | 0.0 |
| Suriname | 10.8 | 89.2 | 85.4 | 77.6 | 51.5 | 7.1 | 87.0 | 13.0 | 3.7 | 1.1 | 0.1 | 0.0 |
| Swaziland | 0.0 | 100.0 | 76.3 | 33.0 | 9.8 | 0.0 | 0.0 | 100.0 | 64.5 | 8.3 | 0.3 | 0.0 |
| Sweden | 0.0 | 100.0 | 99.9 | 96.7 | 62.0 | 25.7 | 5.6 | 94.4 | 60.3 | 21.4 | 1.6 | 0.2 |
| Switzerland | 0.0 | 100.0 | 100.0 | 96.9 | 34.0 | 0.0 | 0.0 | 100.0 | 100.0 | 57.9 | 2.6 | 0.0 |
| Syria | 0.4 | 99.6 | 91.2 | 68.5 | 21.5 | 1.3 | 23.3 | 76.7 | 38.2 | 12.4 | 0.4 | 0.0 |
| Tajikistan | 3.8 | 96.2 | 87.8 | 55.9 | 14.1 | 0.0 | 57.5 | 42.5 | 19.4 | 3.1 | 0.1 | 0.0 |
| Tanzania | 62.7 | 37.3 | 23.0 | 15.4 | 8.8 | 0.0 | 92.5 | 7.5 | 1.3 | 0.2 | 0.0 | 0.0 |
| Thailand | 0.2 | 99.8 | 92.5 | 61.5 | 32.3 | 16.3 | 4.4 | 95.6 | 64.2 | 18.0 | 2.4 | 0.3 |
| The Gambia | 36.1 | 63.9 | 52.8 | 43.6 | 0.0 | 0.0 | 77.2 | 22.8 | 6.7 | 2.6 | 0.0 | 0.0 |
| Timor-Leste | 30.7 | 69.3 | 30.8 | 15.2 | 4.4 | 0.0 | 57.3 | 42.7 | 6.7 | 1.0 | 0.1 | 0.0 |
| Togo | 35.8 | 64.2 | 48.0 | 30.8 | 22.7 | 0.0 | 76.7 | 23.3 | 6.2 | 1.4 | 0.3 | 0.0 |
| Tokelau* | 100.0 | 0.0 | 0.0 | 0.0 | 0.0 | 0.0 | 100.0 | 0.0 | 0.0 | 0.0 | 0.0 | 0.0 |
| Tonga | 0.0 | 100.0 | 98.4 | 66.5 | 0.0 | 0.0 | 0.0 | 100.0 | 79.9 | 13.0 | 0.0 | 0.0 |
| Trinidad and Tobago | 0.0 | 100.0 | 100.0 | 100.0 | 94.3 | 50.2 | 0.0 | 100.0 | 100.0 | 96.6 | 43.5 | 5.2 |
| Tunisia | 0.0 | 100.0 | 99.4 | 80.4 | 48.5 | 16.5 | 9.8 | 90.2 | 61.6 | 17.2 | 1.8 | 0.2 |
| Turkey | 0.0 | 100.0 | 97.8 | 77.7 | 49.9 | 24.3 | 0.0 | 100.0 | 87.4 | 25.7 | 2.2 | 0.3 |
| Turkmenistan | 1.3 | 98.7 | 95.9 | 87.5 | 47.3 | 19.5 | 46.0 | 54.0 | 25.0 | 8.5 | 1.3 | 0.2 |
| Uganda | 60.6 | 39.4 | 17.7 | 9.8 | 4.1 | 0.0 | 83.3 | 16.7 | 3.2 | 0.7 | 0.0 | 0.0 |
| Ukraine | 0.1 | 99.9 | 91.3 | 65.0 | 29.9 | 2.9 | 0.4 | 99.6 | 62.8 | 11.1 | 0.9 | 0.1 |
| United Arab Emirates | 0.0 | 100.0 | 100.0 | 100.0 | 99.3 | 92.7 | 0.0 | 100.0 | 92.3 | 60.9 | 23.4 | 5.7 |
| United Kingdom | 0.0 | 100.0 | 99.9 | 98.2 | 77.0 | 26.0 | 3.6 | 96.4 | 86.4 | 60.8 | 13.5 | 1.4 |
| United States | 0.0 | 100.0 | 99.7 | 97.2 | 77.6 | 36.9 | 30.4 | 69.6 | 46.9 | 23.2 | 3.6 | 0.6 |
| Uruguay | 1.3 | 98.7 | 94.6 | 89.1 | 75.3 | 34.8 | 19.8 | 80.2 | 27.9 | 6.7 | 0.9 | 0.1 |
| Uzbekistan | 0.9 | 99.1 | 96.5 | 81.0 | 19.7 | 2.7 | 56.0 | 44.0 | 25.9 | 10.7 | 0.6 | 0.1 |
| Vanuatu | 41.1 | 58.9 | 40.3 | 14.0 | 0.0 | 0.0 | 69.0 | 31.0 | 11.5 | 7.9 | 1.4 | 0.2 |
| Venezuela | 1.2 | 98.8 | 96.7 | 91.2 | 73.0 | 33.7 | 52.5 | 47.5 | 30.7 | 14.0 | 3.3 | 0.8 |
| Vietnam | 3.2 | 96.8 | 85.6 | 60.6 | 25.2 | 8.2 | 21.3 | 78.7 | 42.5 | 17.5 | 3.0 | 0.7 |
| Virgin Islands* | 0.0 | 100.0 | 100.0 | 99.9 | 75.9 | 0.0 | 0.0 | 100.0 | 100.0 | 99.1 | 37.2 | 0.0 |
| Wallis and Futuna* | 100.0 | 0.0 | 0.0 | 0.0 | 0.0 | 0.0 | 100.0 | 0.0 | 0.0 | 0.0 | 0.0 | 0.0 |
| continued on next page | | | | | | | | | | | | |

| Country | Brightness (μcd/m²) | | | | | | | | | | | |
|---|---|---|---|---|---|---|---|---|---|---|---|---|
| | ≤1.7 | >1.7 | >14 | >87 | >688 | >3000 | ≤1.7 | >1.7 | >14 | >87 | >688 | >3000 |
| | Population (%) | | | | | | Area (%) | | | | | |
| West Bank* | 0.0 | 100.0 | 100.0 | 100.0 | 94.6 | 42.4 | 0.0 | 100.0 | 100.0 | 100.0 | 61.1 | 4.1 |
| Western Sahara* | 4.1 | 95.9 | 95.6 | 95.5 | 90.6 | 5.9 | 93.7 | 6.3 | 1.5 | 0.4 | 0.1 | 0.0 |
| Yemen | 2.7 | 97.3 | 52.5 | 27.9 | 14.4 | 1.4 | 45.5 | 54.5 | 18.9 | 4.9 | 0.5 | 0.1 |
| Zambia | 43.8 | 56.2 | 42.2 | 35.1 | 16.1 | 0.0 | 85.9 | 14.1 | 3.7 | 0.8 | 0.1 | 0.0 |
| Zimbabwe | 45.6 | 54.4 | 37.5 | 28.4 | 2.5 | 0.0 | 77.2 | 22.8 | 4.3 | 0.7 | 0.0 | 0.0 |
| European Union | 0.0 | 100.0 | 99.8 | 94.0 | 59.5 | 20.5 | 1.3 | 98.7 | 88.4 | 48.7 | 6.0 | 0.6 |
| World | 8.0 | 92.0 | 83.2 | 63.7 | 35.9 | 13.9 | 60.3 | 39.7 | 22.5 | 8.6 | 1.2 | 2 |

*Nonindependent territories.

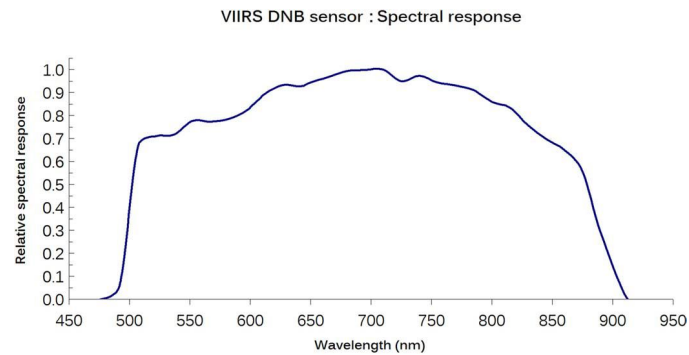

**Fig. 15. VIIRS DNB sensitivity.**

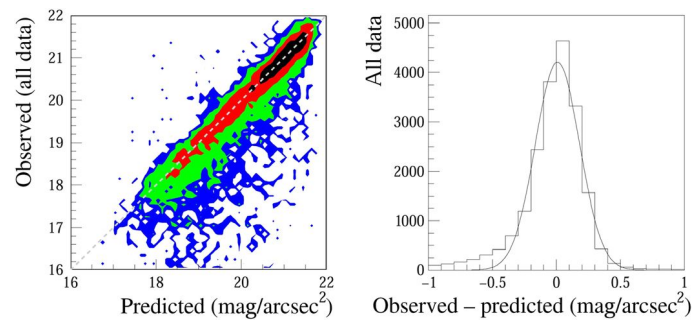

**Fig. 16. Comparisons between sky brightness observations and atlas predictions.** (**Left**) Contour plot comparing the weighted number of SQM observations to the predictions of the atlas in 0.1 mag$_{SQM}$/arcsec² bins. The colors are scaled logarithmically relative to the peak: <0.7 weighted observations (blue), 0.7 to 5 weighted observations (green), 5 to 35.5 weighted observations (red), and 35.5 to 251 weighted observations (black). (**Right**) Fit residuals (observed minus predicted SQM values) for the complete data set, with a Gaussian fit. Negative values mean that the observer reported a sky brighter than the atlas predicted, whereas positive values mean that the observer reported a sky darker than predicted. The tail of brighter-than-predicted observations is usually either contaminated by nearby light sources or taken in different conditions than assumed (for example, unreported clouds, haze, or fog). The histogram shows the total number of observations, not weighted observations, so some locations contribute to multiple entries.

brightness estimated by Duriscoe's model; $S$ is a scaling factor that accounts for the difference in SQM radiance compared to luminance; $W_a$, $W_b$, and $W_c$ are fit weights for the maps; $A$, $B$, and $C$ are the input map predictions for zenith sky luminance; $d$ is a factor that characterizes the change in artificial light as the night goes on; and $h$ is the time in hours after midnight (negative for times before midnight). The parameter $d$ is important because most light-polluted areas display a gradual decrease as the night goes on (45), resulting in differences of around 0.3 mag$_{SQM}$/arcsec² from 20:00 to 02:00. This luminance was converted into radiance (mag$_{SQM}$/arcsec²), and the best fit to the data was obtained by minimizing a likelihood function. The likelihood function assumed an 80% chance that the differences in individual

RESEARCH ARTICLE

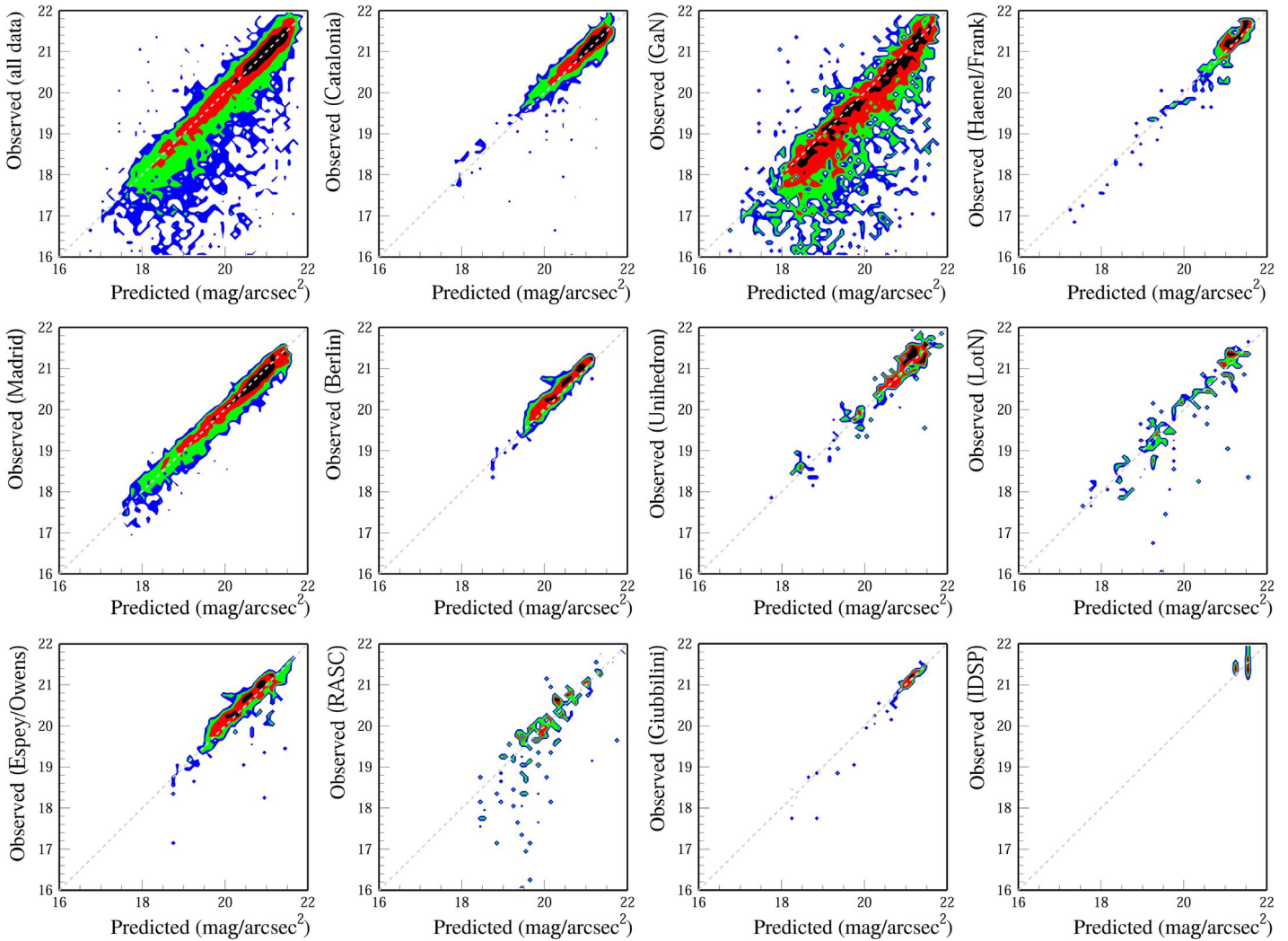

**Fig. 17. Contour plots comparing the weighted number of SQM observations to the predictions of the atlas.** The color scale is logarithmic and scaled relative to the peak value. Observations by citizen scientists tend to be slightly brighter than those made by professionals, with larger tails at very bright values, likely from observations taken too near lamps or under nonideal atmospheric conditions. GaN, Globe at Night; LotN, Loss of the Night app; RASC, Royal Astronomical Society of Canada; IDSP, International Dark Sky Places.
observations from the prediction were drawn from a normal distribution (with a standard deviation of σ) and a 20% chance that the observation was an outlier. The six fit parameters were $S$, $W_a$, $W_b$, $W_c$, $d$, and σ. The weighting factors $W_a$, $W_b$, and $W_c$ were allowed to assume negative values, provided that the light intensity of the resulting upward emission function was positive at all angles.

The best-fit parameters for the global data set were $S = 1.15$, $W_a = 1.9 \times 10^{-3}$, $W_b = 5.2 \times 10^{-4}$, $W_c = 7.6 \times 10^{-5}$, $d = -4.5\%$ per hour, and $\sigma = 0.15$ $mag_{SQM}/arcsec^2$. The angular distribution for upward-directed light with these parameters is shown in red in Fig. 1. Fits were also performed individually for subsets of the data to examine the variation in different regions. In all cases, the map with the largest weight was the map corresponding to Lambertian emission. In most individual cases, the map with peak emission at intermediate angles fit to a negative value. The difference could be related to local orthography (shadowing), which was not taken into account by the radiative transfer model, and to different atmospheric conditions than assumed in the model. This will be examined in more detail in a forthcoming paper.

The comparison between sky brightness and atlas predictions is shown for the complete data set in Fig. 16. The left panel shows a two-dimensional histogram, which was defined with 0.1 $mag_{SQM}/arcsec^2$ bins corresponding to the observed and predicted sky radiances, and filled with the number of weighted observations observed in each bin. The colors are arranged on a logarithmic scale by the number of weighted observations, with blue representing fewer than 1 weighted observation and with red representing between 5 and 35 weighted observations. The outliers were nearly always cases in which the observed sky brightness was markedly larger than predicted by the atlas. Examination of a selection of these locations revealed that they were generally due to observations performed too close to a light source, such as a street lamp. Their effect on the fit was small, both because they represented a small fraction of the total data set and because the likelihood function included an outlier term to account for them. The weight of outlier points was nearly always below 1 because they tended to be clustered inside a city, with other observations included in the same position bin. The histogram on the right-hand panel of Fig. 16 shows the residuals for the entire data set (that is, total number of observations, not weighted observations).

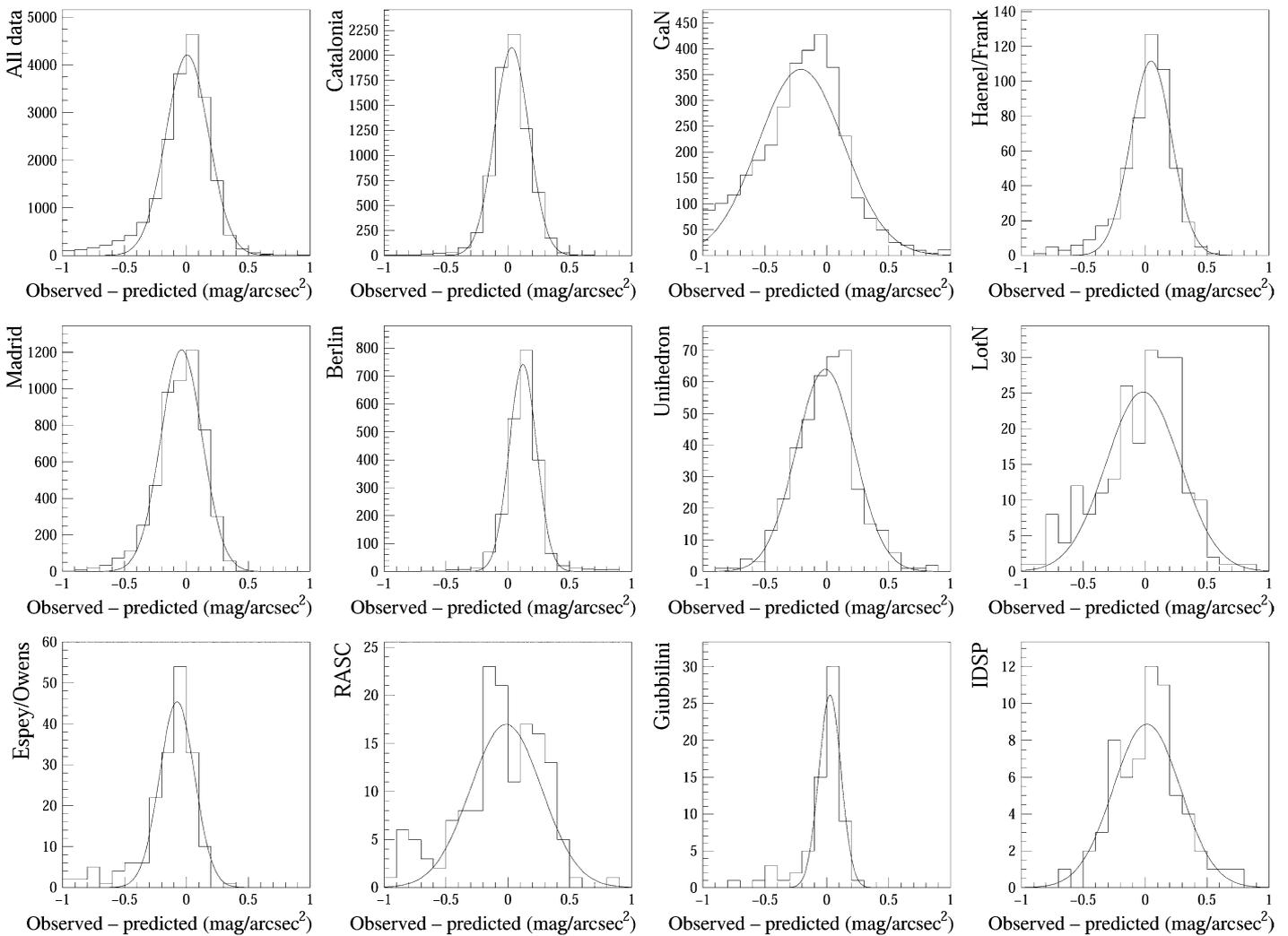

**Fig. 18. Histograms showing the residuals (observed minus predicted SQM values) for the whole data set and for each of the different data providers.** Negative values mean that the observer reported a sky brighter than the atlas predicted, whereas positive values mean that the observer reported a sky darker than predicted. The histograms show the total number of observations, not weighted observations, so some locations contribute to multiple entries. Superimposed are the best-fitting Gaussians.

The global fit does a reasonable job of fitting all individual data sets (Figs. 17, 18, and 19 and Table 3). The color scale in Fig. 17 is logarithmic but scaled to match the size of each individual data set. The data in Figs. 17 and 18 are arranged in the order of the size of the data set. The Catalonia data were taken in driving surveys by professional scientists in Spain. Globe at Night, Unihedron, and Loss of the Night data were taken by citizen scientists worldwide. Haenel/Frank data were taken in Europe and North America, and in particular, include a large number of observations in International Dark Sky Parks and Reserves. Madrid (40) and Berlin data were taken in driving surveys in the respective cities and surroundings by professional scientists. Espey/Owens data were taken by professional and citizen scientists in Ireland. Royal Astronomical Society of Canada data were taken in Canada by volunteers of the Royal Astronomical Society of Canada. Giubbilini data were taken in Italy by a professional scientist. International Dark Sky Places data were taken at a small number of locations in two International Dark Sky Places.

The citizen scientist data include more outliers compared with the data taken by professional scientists. However, these data provide a great value because they include observations in far more diverse settings, including locations outside Europe and the United States, and also in particularly brightly lit areas. The citizen science data also have a broader residual distribution. This is likely partly due to observations being made under a greater range of atmospheric conditions. The Berlin data, for example, have a very narrow distribution, and their mean is darker than the prediction. This is likely because these data were taken on only three nights with exceptional atmospheric clarity.

Consistency checks of the SQM-based calibration were also made using CCD data taken by Falchi (13) and the U.S. National Park Service (46), showing a difference between SQM-calibrated atlas predictions of about 1/10th of a magnitude (Fig. 19, top, and Table 3). The method was also checked by computing maps of the western United States with a higher transparency to better match the transparency conditions of that region, with an aerosol clarity of $K = 0.5$ corresponding to a vertical extinction at sea level of $\Delta m = 0.23$ mag, a horizontal visibility of $\Delta x = 48$ km, and an optical depth of $\tau = 0.21$. At 1000-m altitude, the vertical extinction at sea level becomes $\Delta m = 0.17$ mag, whereas at 2000-m altitude, $\Delta m = 0.13$ mag. The fits with the observed data are shown in Fig. 19 (bottom), but these maps are not presented here. Finally, the fit was compared to a data set of clear sky averages from permanently installed SQM stations (*45*). In this last case, the SD is considerably larger than that for the other data sets and is likely due to a less stringent selection of transparent skies.

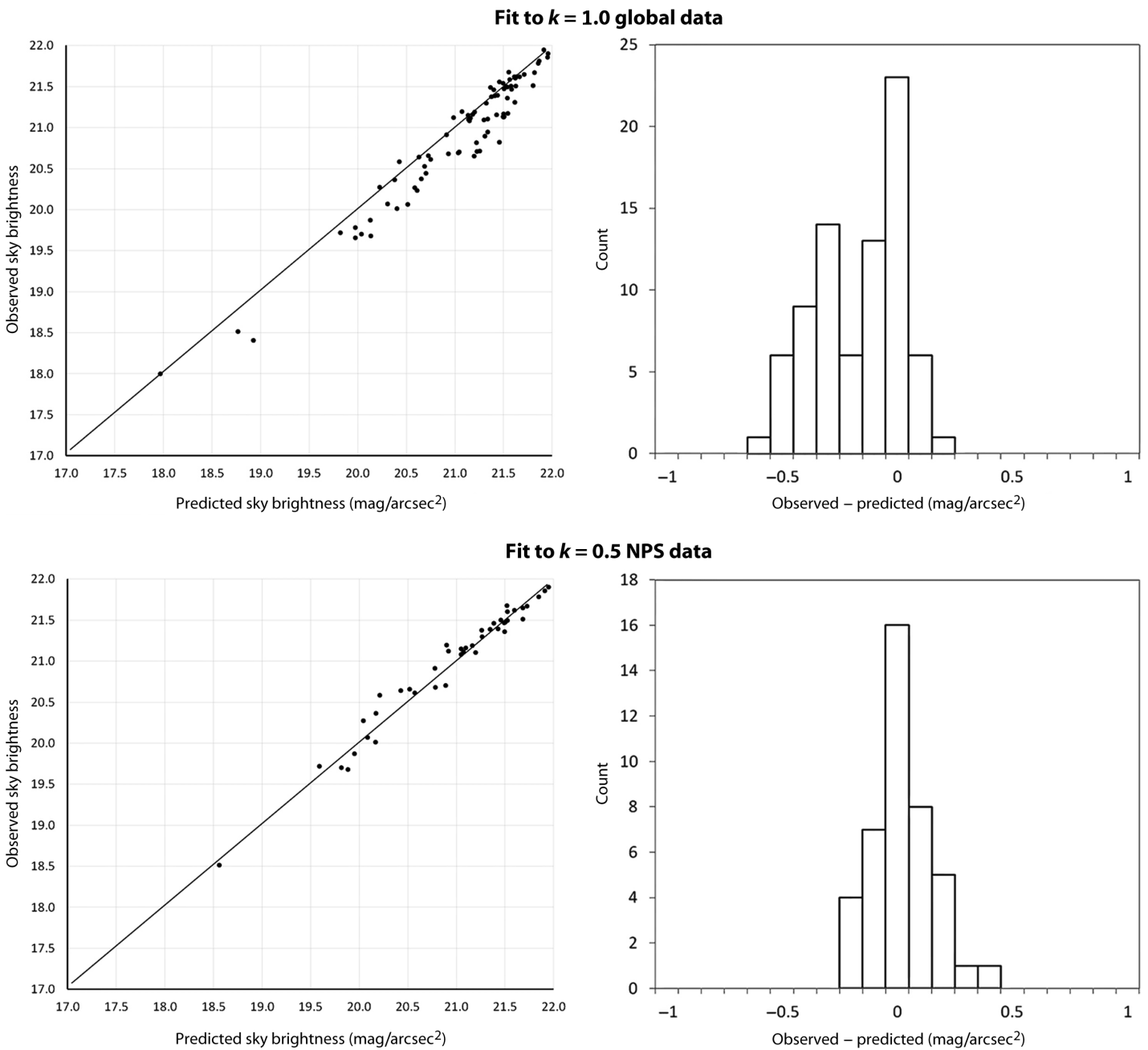

**Fig. 19. Comparisons between sky brightness NPS CCD observations and atlas predictions.** (**Left**) Plots comparing CCD observations to the predictions of the atlas (upper graph, aerosol clarity $K = 1$; lower graph, $K = 0.5$). (**Right**) Fit residuals (observed minus predicted values) for the SQM-based calibration versus CCD NPS observations (upper histogram, $K = 1$; lower histogram, $K = 0.5$).

### Statistics of nighttime brightness

We calculated the percentage of people living under different levels of sky brightness in different countries and the percentage of these countries' areas exposed to different levels of sky brightness using the following sky brightness intervals:

Table 3. **Performance of atlas predictions for different data sets.** Negative values indicate that the observed skies were brighter than predicted. The top entries are used to calibrate the atlas. The final three entries are independent data sets used for verification. PS, professional scientist; CS, citizen scientist.

| Provider | Location | Mode | Mean | σ | Scientist |
|---|---|---|---|---|---|
| All data | Worldwide | Both | 0.008 | 0.17 | Mixed |
| Various | Catalonia, Spain | Car | 0.03 | 0.14 | PS |
| Universidad Complutense de Madrid | Madrid, Spain | Car | −0.04 | 0.17 | PS |
| Kyba/Sanchez | Berlin, Germany | Car | 0.12 | 0.11 | PS |
| Globe at Night | Worldwide | Hand | −0.21 | 0.34 | CS |
| Haenel/Frank | Europe and North America | Hand | 0.05 | 0.17 | PS |
| Unihedron | Worldwide | Hand | −0.01 | 0.24 | CS |
| Loss of the Night app | Worldwide | Hand | −0.02 | 0.29 | CS |
| Espey/Owens | Ireland | Hand | −0.08 | 0.14 | Mixed |
| Royal Astronomical Society of Canada | Canada | Hand | −0.01 | 0.28 | CS |
| Giubbilini | Italy | Hand | 0.03 | 0.09 | PS |
| International Dark Sky Places | Wales | Hand | 0.01 | 0.27 | CS |
| U.S. National Park Service | United States | CCD | 0.11 | 0.13 | PS |
| Falchi | Italy | CCD | −0.04 | 0.19 | PS |
| SQM stations | Worldwide | Permanent stations | −0.06 | 0.46 | Mixed |

(i) Up to 1% above the natural light (0 to 1.7 μcd/m$^2$)—pristine sky

(ii) From 1 to 8% above the natural light (1.7 to 14 μcd/m$^2$)—relatively unpolluted at the zenith but degraded toward the horizon

(iii) From 8 to 50% above natural nighttime brightness (14 to 87 μcd/m$^2$)—polluted sky degraded to the zenith

(iv) From 50% above natural to the level of light under which the Milky Way is no longer visible (87 to 688 μcd/m$^2$)—natural appearance of the sky is lost

(v) From Milky Way loss to estimated cone stimulation (688 to 3000 μcd/m$^2$)

(vi) Very high nighttime light intensities (>3000 μcd/m$^2$)—night adaptation is no longer possible for human eyes

First, we converted the world country polygons, obtained from the ESRI Data and Maps Media Kit (47), into a raster file of 30–arcsec resolution. Next, we linked each pixel in this file to the pixel values obtained from two other raster files under analysis—our raster file of artificial zenith sky brightness maps and the raster file of global population (48). The data merging was performed with the ArcGIS 10.x software, using its "extract multiple values" raster-processing feature. Next, we used the SPSS version 22 statistical software to aggregate the population counts and land area shares into different exposure groups corresponding to the aforementioned levels of nighttime light brightness.

## REFERENCES AND NOTES

1. P. Cinzano, F. Falchi, C. D. Elvidge, K. E. Baugh, The artificial night sky brightness mapped from DMSP satellite Operational Linescan System measurements. *Mon. Not. R. Astron. Soc.* **318**, 641–657 (2000).
2. R. H. Garstang, Night-sky brightness at observatories and sites. *Publ. Astron. Soc. Pac.* **101**, 306–329 (1989).
3. M. Smith, Time to turn off the lights. *Nature* **457**, 27 (2009).
4. K. J. Gaston, M. E. Visser, F. Hölker, The biological impacts of artificial light at night: The research challenge. *Philos. Trans. R. Soc. London Ser. B* **370**, 20140133 (2015).
5. P. Cinzano, F. Falchi, C. D. Elvidge, The first World Atlas of the artificial night sky brightness. *Mon. Not. R. Astron. Soc.* **328**, 689–707 (2001).
6. D. M. Duriscoe, C. B. Luginbuhl, C. A. Moore, Measuring night-sky brightness with a wide-field CCD camera. *Publ. Astron. Soc. Pac.* **119**, 192–213 (2007).
7. P. Cinzano, F. J. D. Castro, The artificial sky luminance and the emission angles of the upward light flux. *Mem. Soc. Astron. Ital.* **71**, 251 (2000).
8. C. B. Luginbuhl, C. E. Walker, R. J. Wainscoat, Lighting and astronomy. *Phys. Today* **62**, 32 (2009).
9. F. G. Smith, Report and recommendations of IAU Commission 50, Reports on astronomy. *IAU Trans.* **XVIIA**, 218 (1979).
10. Commission Internationale de l'Eclairage, *Light as a True Visual Quantity: Principles of Measurement*. Publication CIE No. 41 (ed. 1, 1978; reprint, 1994) (Bureau Central de la CIE, Paris, 1994).
11. F. Patat, O. S. Ugolnikov, O. V. Postylyakov, UBVRI twilight sky brightness at ESO-Paranal. *Astron. Astrophys.* **455**, 385–393 (2006).
12. M. Aubé, Physical behaviour of anthropogenic light propagation into the nocturnal environment. *Philos. Trans. R. Soc. London Ser. B* **370**, 20140117 (2015).
13. F. Falchi, Campaign of sky brightness and extinction measurements using a portable CCD camera. *Mon. Not. R. Astron. Soc.* **412**, 33–48 (2010).
14. P. Cinzano, F. Falchi, C. D. Elvidge, Recent progresses on a Second World Atlas of the night-sky brightness, in *Starlight—A Common Heritage*, C. Marin, J. Jafari, Eds. (Starlight Initiative and Instituto de Astrofisica de Canarias, Canary Islands, Spain, 2007), pp. 385–400.
15. C. C. M. Kyba, T. Ruhtz, J. Fischer, F. Hölker, Cloud coverage acts as an amplifier for ecological light pollution in urban ecosystems. *PLOS One* **6**, e17307 (2011).
16. F. Falchi, P. Cinzano, C. D. Elvidge, D. M. Keith, A. Haim, Limiting the impact of light pollution on human health, environment and stellar visibility. *J. Environ. Manage.* **92**, 2714–2722 (2011).
17. M. Aubé, J. Roby, M. Kocifaj, Evaluating potential spectral impacts of various artificial lights on melatonin suppression, photosynthesis, and star visibility. *PLOS One* **8**, e67798 (2013).
18. C. B. Luginbuhl, P. A. Boley, D. R. Davis, The impact of light source spectral power distribution on sky glow. *J. Quant. Spectrosc. Radiat. Transfer* **139**, 21–26 (2014).
19. H. Jin, S. Jin, L. Chen, S. Cen, K. Yuan, Research on the lighting performance of LED street lights with different color temperatures. *IEEE Photonics J.* **7**, 1–9 (2015).
20. G. Tosini, I. Ferguson, K. Tsubota, Effects of blue light on the circadian system and eye physiology. *Mol. Vis.* **22**, 61–72 (2016).

21. C. Rich, T. Longcore, *Ecological Consequences of Artificial Night Lighting* (Island Press, Washington, DC, 2005).
22. S. M. Pauley, Lighting for the human circadian clock: Recent research indicates that lighting has become a public health issue. *Med. Hypotheses* **63**, 588–596 (2004).
23. R. G. Stevens, D. E. Blask, G. C. Brainard, J. Hansen, S. W. Lockley, I. Provencio, M. S. Rea, L. Reinlib, Meeting report: The role of environmental lighting and circadian disruption in cancer and other diseases. *Environ. Health Perspect.* **115**, 1357–1362 (2007).
24. A. Haim, B. A. Portnov, *Light Pollution as a New Risk Factor for Human Breast and Prostate Cancers* (Springer, Dordrecht, 2013).
25. T. Gallaway, R. N. Olsen, D. M. Mitchell, The economics of global light pollution. *Ecol. Econ.* **69**, 658–665 (2010).
26. P. R. Marchant, A demonstration that the claim that brighter lighting reduces crime is unfounded. *Br. J. Criminol.* **44**, 441–447 (2004).
27. R. Steinbach, C. Perkins, L. Tompson, S. Johnson, B. Armstrong, J. Green, C. Grundy, P. Wilkinson, P. Edwards, The effect of reduced street lighting on road casualties and crime in England and Wales: Controlled interrupted time series analysis. *J. Epidemiol. Community Health* **69**, 1118–1124 (2015).
28. P. Cinzano, C. D. Elvidge, Night sky brightness at sites from DMSP-OLS satellite measurements. *Mon. Not. R. Astron. Soc.* **353**, 1107–1116 (2004).
29. M. Aubé, M. Kocifaj, Using two light-pollution models to investigate artificial sky radiances at Canary Islands observatories. *Mon. Not. R. Astron. Soc.* **422**, 819–830 (2012).
30. D. M. Duriscoe, C. B. Luginbuhl, C. D. Elvidge, The relation of outdoor lighting characteristics to sky glow from distant cities. *Light. Res. Technol.* **46**, 35–49 (2014).
31. F. Falchi, Luminanza artificiale del cielo notturno in Italia, Master's thesis, Università di Milano (1998).
32. F. Falchi, P. Cinzano, Maps of artificial sky brightness and upward emission in Italy from DMSP satellite measurements. *Mem. Soc. Astron. Ital.* **71**, 139 (2000).
33. C. D. Elvidge, K. E. Baugh, M. Zhizhin, F.-C. Hsu, Why VIIRS data are superior to DMSP for mapping nighttime lights. *Proc. Asia-Pac. Adv. Network* **35**, 62 (2013).
34. K. Baugh, F.-C. Hsu, C. D. Elvidge, M. Zhizhin, Nighttime lights compositing using the VIIRS day-night band: Preliminary results. *Proc. Asia-Pac. Adv. Network* **35**, 70 (2013).
35. F. Falchi, P. Cinzano, D. Duriscoe, C. C. M. Kyba, C. D. Elvidge, K. Baugh, B. Portnov, N. A. Rybnikova, R. Furgoni, Supplement to the New World Atlas of Artificial Night Sky Brightness, GFZ Data Services (2016); http://doi.org/10.5880/GFZ.1.4.2016.001.
36. P. Cinzano, F. Falchi, C. D. Elvidge, Naked-eye star visibility and limiting magnitude mapped from DMSP-OLS satellite data. *Mon. Not. R. Astron. Soc.* **323**, 34–46 (2001).
37. R. H. Garstang, Model for artificial night-sky illumination. *Publ. Astron. Soc. Pac.* **98**, 364–375 (1986).
38. D. B. Gesch, K. L. Verdin, S. K. Greenlee, New land surface digital elevation model covers the Earth. *Eos Trans. Am. Geophys. Union* **80**, 69–70 (1999).
39. M. Kocifaj, T. Posch, H. A. Solano Lamphar, On the relation between zenith sky brightness and horizontal illuminance. *Mon. Not. R. Astron. Soc.* **446**, 2895–2901 (2015).
40. J. Zamorano, A. S. de Miguel, F. Ocaña, B. Pila-Díez, J. Gómez Castaño, S. Pascual, C. Tapia, J. Gallego, A. Fernández, M. Nievas, Testing sky brightness models against radial dependency: A dense two dimensional survey around the city of Madrid, Spain. *J. Quant. Spectrosc. Radiat. Transfer* 10.1016/j.jqsrt.2016.02.029 (2016).
41. C. C. M. Kyba, J. M. Wagner, H. U. Kuechly, C. E. Walker, C. D. Elvidge, F. Falchi, T. Ruhtz, J. Fischer, F. Hölker, Citizen science provides valuable data for monitoring global night sky luminance. *Sci. Rep.* **3**, 1835 (2013).
42. D. M. Duriscoe, Measuring anthropogenic sky glow using a natural sky brightness model. *Publ. Astron. Soc. Pac.* **125**, 1370–1382 (2013).
43. K. Krisciunas, D. R. Semler, J. Richards, H. E. Schwarz, N. B. Suntzeff, S. Vera, P. Sanhueza, Optical sky brightness at Cerro Tololo Inter-American observatory from 1992 to 2006. *Publ. Astron. Soc. Pac.* **119**, 687–696 (2007).
44. P. Cinzano, F. Falchi, The propagation of light pollution in the atmosphere. *Mon. Not. R. Astron. Soc.* **427**, 3337–3357 (2012).
45. C. C. M. Kyba, K. P. Tong, J. Bennie, I. Birriel, J. J. Birriel, A. Cool, A. Danielsen, T. W. Davies, P. N. den Outer, W. Edwards, R. Ehlert, F. Falchi, J. Fischer, A. Giacomelli, F. Giubbilini, M. Haaima, C. Hesse, G. Heygster, F. Hölker, R. Inger, L. J. Jensen, H. U. Kuechly, J. Kuehn, P. Langill, D. E. Lolkema, M. Nagy, M. Nievas, N. Ochi, E. Popow, T. Posch, J. Puschnig, T. Ruhtz, W. Schmidt, R. Schwarz, A. Schwope, H. Spoelstra, A. Tekatch, M. Trueblood, C. E. Walker, M. Weber, D. L. Welch, J. Zamorano, K. J. Gaston, Worldwide variations in artificial skyglow. *Sci. Rep.* **5**, 8409 (2015).
46. U.S. National Park Service, *Night Sky Monitoring Database*, http://nature.nps.gov/night/skymap.cfm [accessed May 28, 2016].
47. ArcGIS, *World Countries*, http://www.arcgis.com/home/item.html?id=3864c63872d84ae-c91933618e3815dd2 [accessed May 28, 2016].
48. Oak Ridge National Laboratory, *LandScan*™, http://web.ornl.gov/sci/landscan/ [accessed May 28, 2016].

**Acknowledgments:** We are grateful to the individuals and groups who provided the sky brightness data: J. Zamorano, A. Sanchez de Miguel, S. J. Ribas, A. Haenel, S. Frank, F. Giubbilini, B. Espey, S. Owens, Parc Astronòmic Montsec, Dir. Gral de Qualitat Ambiental de la Generalitat de Catalunya, Institut d'Estudis Espaials de Catalunya (El Instituto de Ciencias del Cosmos–Universitat de Barcelona), Attivarti.org, the Royal Astronomical Society of Canada, the International Dark Sky Association, and hundreds of anonymous citizen scientists. We thank the European Cooperation in Science and Technology (COST) Action Loss of the Night Network (ES 1204) for making it possible for several authors to meet to discuss the work in person. In addition, much of the Berlin data was acquired during a COST-funded short-term scientific mission. We thank H. Kuechly for extracting the map predictions for each of the SQM locations. F.F. is indebted to M. G. Smith, P. Sanhueza, C. Marin, and C. R. Smith for their support during the years of gestation of this project. F.F. thanks S. Klett, A. Weekes of iCandi Apps Ltd., A. Crumey, A. B. Watson, and A. J. Zele for their contributions at different levels. **Funding:** Part of the preliminary research carried out at Istituto di Scienza e Tecnologia dell'Inquinamento Luminoso was supported by the Italian Space Agency (contract I/R/160/02). No specific funds were used for this work. **Author contributions:** F.F. led and designed the study, wrote the manuscript, contributed to the sky brightness CCD data, analyzed the statistics, and performed the software computation. P.C. developed the light pollution propagation model and wrote the software to compute sky brightness. C.C.M.K. gathered the SQM data and calibrated the maps using them and other data sets. D.D. contributed to the sky brightness calibration, assembled and produced the final maps, and led the collection of U.S. National Park Service CCD brightness data. C.D.E. and K.B. collected and assembled the nighttime satellite upward radiance data. B.A.P. and N.A.R. performed the statistical computation. R.F. performed the software computation. F.F. and R.F. performed statistical analysis on population and area data. C.C.M.K., D.D., K.B., and B.A.P. wrote parts of the manuscript. F.F., D.D., C.C.M.K., and R.F. produced the figures. All authors read and commented on the manuscript. **Competing interests:** The authors declare that they have no competing interests. **Data and materials availability:** All data needed to evaluate the conclusions in the paper are present in the paper. Additional data of night sky brightness measurements related to this paper may be requested from the corresponding author or from http://doi.org/10.5880/GFZ.1.4.2016.001 (35) or from the following: F. Giubbilini data: Buiometria Partecipativa project (buiometria@attivarti.org); RASC data: http://old.rasc.ca/~admin/sqm/SQM_data_view.php; Catalonia data: darksky@parcastronomic.cat; Espey/Owens data: Brian.Espey@tcd.ie and steve@owens-online.co.uk; Haenel/Frank data: ahaenel@uos.de; Madrid data: http://dx.doi.org/10.5281/zenodo.51713; Globe at Night data: http://www.globeatnight.org/; Unihedron data: http://www.unihedron.com/projects/darksky/database/; and IDSP data: J. Barentine (john@darksky.org).

Submitted 22 February 2016
Accepted 20 May 2016
Published 10 June 2016
10.1126/sciadv.1600377

**Citation:** F. Falchi, P. Cinzano, D. Duriscoe, C. C. M. Kyba, C. D. Elvidge, K. Baugh, B. A. Portnov, N. A. Rybnikova, R. Furgoni, The new world atlas of artificial night sky brightness. *Sci. Adv.* **2**, e1600377 (2016).